\documentclass[fleqn,usenatbib]{mnras}

\usepackage{newtxtext,newtxmath}
\usepackage[T1]{fontenc}

\DeclareRobustCommand{\VAN}[3]{#2}
\let\VANthebibliography\thebibliography
\def\thebibliography{\DeclareRobustCommand{\VAN}[3]{##3}\VANthebibliography}

\usepackage{graphicx}	% Including figure files
\usepackage{amsmath}	% Advanced maths commands
\usepackage{lscape}
\usepackage{makecell}

\title[Massive spirals observed through UVIT]{Probing star formation in five of the most massive spiral galaxies observed through {\it ASTROSAT UltraViolet Imaging Telescope} }

\author[Ray et al.]{
Shankar Ray,$^{1,2}$\thanks{E-mail: shankarray.physics@gmail.com}
Suraj Dhiwar,$^{3,4}$
Joydeep Bagchi,$^{1}$
and M. B. Pandge$^{2}$\thanks{E-mail: mbpandge@gmail.com}
\\
$^{1}$Department of Physics and Electronics, Christ University, Hosur Road, Bengaluru 560029, India\\
$^{2}$Department of Physics and Electronics, Dayanand Science College, Latur 413512, India\\
$^{3}$Inter-University Centre for Astronomy and Astrophysics, Pune 411007, India\\
$^{4}$Savitribai Phule Pune University, Pune 411007, India\\
}

\date{Accepted XXX. Received YYY; in original form ZZZ}

\pubyear{2023}

\begin{document}
\label{firstpage}
\pagerange{\pageref{firstpage}--\pageref{lastpage}}
\maketitle

%%%%% AUTHORS - PLACE YOUR OWN COMMANDS HERE %%%%%
\newcommand{\ray}[1]{\textcolor{black}{\bf Ray: #1}}
\newcommand{\mahadev}[1]{\textcolor{red}{\bf Mahadev: #1}}
\newcommand{\suraj}[1]{\textcolor{blue}{\bf Suraj: #1}}
\newcommand{\vdag}{(v)^\dagger}
\newcommand\aastex{AAS\TeX}
\newcommand\latex{La\TeX}
\newcommand{\change}[1]{{\bf \color{red} #1}}
\newcommand\asca{{\it ASCA}}
\newcommand\sax{{\it BeppoSAX}}
\newcommand\chandra{{\it Chandra}}
\newcommand\rosat{{\it ROSAT}}
\newcommand\rxte{{\it RXTE}}
\newcommand\sdss{{\it SDSS}}
\newcommand\xmm{{\it XMM-Newton}}
\newcommand\ftools{{\it FTOOLS}}
\newcommand\s{{\rm~s}}
\newcommand\ks{{\rm~ks}}
\newcommand\mhz{{\rm~mHz}}
\newcommand\mpc{{\rm~Mpc}}
\newcommand\pc{{\rm~pc}}
\newcommand\hz{{\rm~Hz}}
\newcommand\kev{{\rm~keV}}
\newcommand\ev{{\rm~eV}}
\newcommand\kms{\ifmmode {\rm~km\ s}^{-1} \else ~km s$^{-1}$\fi}
\newcommand\Hunit{\ifmmode {\rm~km\ s}^{-1}\ {\rm Mpc}^{-1}
        \else ~km s$^{-1}$ Mpc$^{-1}$\fi}
\newcommand\ctssec{\ifmmode {\rm~count\ s}^{-1} \else ~count s$^{-1}$\fi}
\newcommand\ergsec{\ifmmode {\rm~erg\ s}^{-1} \else
        ~erg s$^{-1}$\fi}
\newcommand\funit{\ifmmode {\rm~erg\ s}^{-1}\;{\rm cm}^{-2} \else
        ~ergs s$^{-1}$ cm$^{-2}$\fi}
\newcommand\phflux{\ifmmode {\rm~photon\ s}^{-1}\;{\rm cm}^{-2}
        \else   ~photon s$^{-1}$ cm$^{-2}$\fi}
\newcommand\efluxA{\ifmmode {\rm~erg\ s}^{-1}\;{\rm cm}^{-2}\;{\rm
        \AA}^{-1} \else ~erg s$^{-1}$ cm$^{-2}$ \AA$^{-1}$\fi}
\newcommand\efluxHz{\ifmmode {\rm~erg\ s}^{-1}\;{\rm cm}^{-2}\;{\rm
        Hz}^{-1} \else ~erg s$^{-1}$ cm$^{-2}$ Hz$^{-1}$\fi}
\newcommand\cc{\ifmmode {\rm~cm}^{-3} \else cm$^{-3}$\fi}
\newcommand\FWHM{\ifmmode {\rm~FWHM} \else ${\rm~FWHM}$\fi}
\newcommand\Zsun{\ifmmode Z_{\odot} \else $M_{\odot}$\fi}
\newcommand\Lsun{\ifmmode L_{\odot} \else $L_{\odot}$\fi}
\newcommand\ltsim{\raisebox{-.5ex}{$\;\stackrel{<}{\sim}\;$}}
\newcommand\gtsim{\raisebox{-.5ex}{$\;\stackrel{>}{\sim}\;$}}
\newcommand\hbeta{\ifmmode {\rm H}\beta \else H$\beta$\fi}
\newcommand\Kalpha{\ifmmode {\rm K}\alpha \else K$\alpha$\fi}
\newcommand\nh{\ifmmode N_{\rm H} \else N$_{\rm H}$\fi}
\newcommand{\Ha}{H$\alpha$ }
\newcommand{\flux}{erg\,cm$^{-2}$\,s$^{-1}$}
\newcommand{\lum}{erg\,s$^{-1}$}
\newcommand{\TwoMASS}{The Two Micron All Sky Survey}
\newcommand{\Teff}{\ensuremath{T_{\mathrm{eff}}}}
\newcommand{\Msun}{\ensuremath{\rm M_{\odot~}}}
%%%%%%%%%%%%%%%%%%%%%%%%%%%%%%%%%%%%%%%%%%%%%%%%%%

\begin{abstract}
We present highly resolved and sensitive imaging of the five nearby massive spiral galaxies (with rotation velocities $\rm > 300 km s^{-1}$) observed by the UltraViolet Imaging Telescope onboard India's  multi-wavelength astronomy satellite ASTROSAT, along with other archival observations. These massive spirals show a far-ultraviolet star formation rate in the range of $\sim$$\rm 1.4$$-$$\rm13.7 M_{\odot} yr^{-1}$ and fall in the `Green Valley' region with a specific star formation rate within $\sim$$\rm10^{-11.5}$$-$$\rm10^{-10.5}yr^{-1}$. Moreover, the mean star formation rate density of the highly resolved star-forming clumps of these objects are in the range $\rm 0.011$$-$$\rm 0.098 M_{\odot} yr^{-1}kpc^{-2}$, signifying localised star formation. From the spectral energy distributions, under the assumption of a delayed star formation model, we show that the star formation of these objects had peaked in the period of $\sim$ $\rm0.8$$-$$\rm2.8$ Gyr after the `Big Bang' and the object that has experienced the peak sooner after the `Big Bang' show relatively less star-forming activity at $\rm\it{z}$$\sim$0 and falls below the main-sequence relation for a stellar content of $\rm \gtsim 10^{11} M_{\odot}$. We also show that these objects accumulated much of their stellar mass in the early period of evolution with $\sim31$$-$$42$ per cent of the total stellar mass obtained in a time of $(1/16)$$-$$(1/5)^{\rm th}$ the age of the Universe. We estimate that these massive objects convert their halo baryons into stars with efficiencies falling between $\sim 7-31$ percent.
%$\rm SFR(t) \propto \frac{t}{\tau^{2}}exp(-t/\tau)$
%\textcolor{red}{These objects also host central supermassive black holes of mass $\rm 10^{7-8}~\rm M_{\odot}$ which arguably experienced rapid growth as of the stellar mass to be able to eventually regulate the star-forming activity in the later period of galaxy evolution. Our study is a significant step forward in understanding complex star formation history in massive galaxies.}

\end{abstract}


%%%%%%%%%%%%%%%%%%%%%%%%%%%%%%%%%%%%%%%%%%%%%%%%%%
%The observed objects are NGC~1030, NGC~1961, NGC~4501, NGC~5635, NGC~266 within the distance between 33 and 125 Mpc%. We have selected this sample to understand 
%These massive and highly rotating galaxies show a mix of moderate to nearly quenched star formation with dust corrected FUV star formation rate in the range, $\rm \sim 1.4-13.7 M_\odot yr^{-1}$ with mean star formation rate density of star forming clumps in the range $\rm \sim 0.01-0.07 M_{\odot} yr^{-1}kpc^{-2}$ signifying highly localised young star formations.%
%These massive and highly rotating galaxies show  moderate to nearly quenched star formation rates. 
%%%%%%%%%%%%%%%%% BODY OF PAPER %%%%%%%%%%%%%%%%%%

\section{Introduction} \label{sec:introduction}
Significant progress over the past few decades has been made in the understanding of the formation of galaxies and their evolution with time. However, a fully developed theory of galaxy formation remains one of the  significant frontier problems of astrophysics (see \citealt{1977ARA&A..15..235G,2006ARA&A..44..193B,2015ARA&A..53...51S,2017ARA&A..55...59N,2022ARA&A..60..121R} for a review of galaxy formation theories, also see \citealt{1978MNRAS.183..341W,1991ApJ...379...52W,2006ApJ...639..590F,2006ApJ...644L...1S,2000MNRAS.319..168C} for some leading works on galaxy formation). Moreover, explaining galaxy formation and evolution across cosmic time involves understanding  processes across many branches of physics, starting from cosmology to plasma physics, which necessarily span a vast range of lengths and timescales. The current galaxy formation scenario is discussed under the hierarchical structure formation model of lambda cold dark matter cosmology $\rm \Lambda$CDM (see \citealt{2010PhR...495...33B} for a review). Specifically, it is still 
uncertain how  and when  the  rare, incredibly massive, rotationally supported spiral galaxies  observed in the local Universe,  with stellar   masses $> 10^{11}$ \Msun and viral masses $> 10^{13}$ \Msun, have  formed, as  it depends on poorly understood non-linear baryonic physics and complex interaction with the dark matter halos. However, based on cosmological simulations, \citet{2010ApJ...725.2312O} propose a `two-phase' evolution of galaxies consisting of a rapid early phase in $\it{z}\rm \gtsim2$ followed by an extended phase of evolution for $\it{z}\rm \ltsim3$. They suggest that progenitor of massive galaxies grew considerably in the extended phase by accreting and merging with smaller satellite galaxies formed even before $\it{z}\rm\sim3$ around the gravitational well of the central galaxy (see \citealt{2012MNRAS.425..641L,2016MNRAS.458.2371R} also). They also emphasise that the massive galaxies have a considerably higher population of older stars, which is a result of merging with systems that already had `in situ' star formation before $\it{z}\rm\sim2$ and they grew in physical dimension according to hierarchical clustering as proposed by $\rm \Lambda$CDM cosmology. \citet{2009ApJ...703..785D} provide a theoretical background of the formation of massive galaxies at high red-shift, where the interplay between smooth, clumpy cold streams, instability in the disc and bulge formation govern the growth and evolution of the early phase of the galaxies.

%They also  predict the emergence of bi-modality in the galaxy type by red-shift $\it{z}\rm\sim3$ with star-forming discs containing giant clumps forming stars and bulge/spheroid-dominated galaxies with already suppressed star-forming activity. 

%Also, a major question in galaxy formation is, while some  nearby  galaxies still are actively forming young stars whereas  others  have shut down their gas-to-star conversions effectively.  Observations show that 

Recent observations have found the existence of massive galaxies at the early Universe. For instance, \citet{2022arXiv220712446L} reported the finding of six massive galaxies with stellar mass $\rm \gtsim 10^{10} M_{\odot}$ at red-shift $\rm 7.4\leq\it{z}\rm\leq9.1$, \citet{2023arXiv230401378H} reported the existence of Ultra Luminous Infrared Galaxies (ULIRG) with infrared luminosity $\rm L_{IR}>10^{12.5} L_{\odot}$ and stellar masses $\rm \sim 10^{11} M_{\odot}$ at red-shift $\rm \it{z}\rm\sim2$, \citet{2023ApJ...942L..19C} found 16 galaxies with stellar mass $\rm > 10^{10.5}M_{\odot}$ at red-shift $\rm 1<\it{z}<\rm 4.5$. All these findings indicate that some of these objects had experienced a period of intense star formation and stellar mass growth in a very short period of time in the early Universe. However, the general trend for galaxies, verified by observations, has been that the star formation rate reached  its  peak at the `cosmic high noon' or `cosmic noon' near  redshift $\it{z} \rm \approx 1 - 2$  when galaxies were rapidly building up their stellar masses by converting  cosmic baryons to stars, and subsequently, over the next  few billions of years the star formation  went down drastically \citep{MadauDickinson2014}.

Eventually, by the  present  era,  most of the massive galaxies are supposed to have  stopped growing \citep[][]{2004MNRAS.351.1151B, 2015ApJ...801L..29R},  having reached a state of apparent quiescence \citep{2010ApJ...717..379B,2019MNRAS.488.3143B,2013MNRAS.428.3121M}.  And, for reasons still  unknown,  in  our  local universe, some  low-mass  galaxies  are still actively forming stars,  while   the majority of  massive  galaxies with halo masses  $\rm M_{halo} \gtsim 10^{12} \Msun$  display a  strikingly  low specific star formation  rate sSFR  (where  sSFR = SFR/stellar mass; SFR being the star formation rate) compared to less massive galaxies (see \citet{2012NewAR..56...93A} and \citet[][]{2018ARA&A..56..435W} for reviews). One such extremely massive, low star formation rate, `red and dead'  spiral galaxy UGC~12591, was reported in \citet{2022MNRAS.517...99R}. 

%Stars in galaxies form in the regions having surrounding temperature of $\rm  \ltsim 10^{2}$K after cooling has taken the place of the accreted gas in the galactic disc, from the hot halo having temperature in the range $\rm \sim 10^{3}-10^{6}$K within the virial radius \citep{2018NatAs...2..695M}. 

The suppression of star formation in these galaxies could be due to multiple factors contributing together and/or functional over different timescales of galaxies' evolution. Further, different quenching mechanisms may dominate at different mass ranges of galaxies \citep{2007ApJS..173..619K,Ciconeetal2014,ForsterSchreiberetal2014,2023MNRAS.518.4943D}. For massive galaxies with halo mass $\rm \gtsim 10^{12}M_{\odot}$, the halo gas gets shocked heated \citep{1977MNRAS.179..541R} while collapsing onto the dark matter halo, which can delay the star-forming activity in the host galaxy's disc \citep{2003MNRAS.345..349B}. For massive galaxies at high red-shift $\it{z}\rm\sim2$, \citet{2015MNRAS.446.1939F} show that the suppression of star formation happens due to reduced accretion of gas from the intergalactic medium to the galactic dark matter halo, termed as `Cosmological Starvation' and suggest the presence of additional mechanism such as radio-mode feedback to maintain the quenched state up to the present times \citep{2006MNRAS.370..645B,2006MNRAS.365...11C}. \citet{2023ApJ...953..119P} found that the massive galaxies at the 'cosmic noon' are formed from a major starburst and are rapidly quenched by AGN feedback. 
%\textbf{Barring the effects of feedback from the active galactic nuclei on star formation of galaxies, it has also been noticed that the stellar mass plays a crucial role in the suppression of star formation in massive galaxies, while the halo mass could be responsible for regulating the quenching process in lower mass galaxies \citep{2019ApJ...878...69L}. 
While, from the morphological analysis of the star-forming regions of galaxies with stellar masses $\rm >10^{11.3}M_{\odot}$, \citet{2020ApJ...895..100X} conclude that one-fifth of the massive galaxies are still forming stars, and overwhelmingly most of them have gone through recent mergers. 

%\citet{2009ApJ...707..250M} propose that the early-type galaxies can be suppressed to form stars, not affected by the existence of gas consumption and removal or termination of gas supply if there is a growth of a central spheroid which stabilises the galactic disc as star formation takes place in gravitationally unstable molecular clouds in the gas disc. However, on the contrary, some recent works have found that some early-type galaxies are actively forming stars \citet{2004ApJ...601L.127F,2007ApJS..173..619K,2009MNRAS.393.1324B,2012MNRAS.421..314C,2013MNRAS.432..359T, 2023MNRAS.518.4943D,2023A&A...671A.166G}. 

Moreover, massive galaxies with rotational velocities $\rm v_{rot}\gtsim300kms^{-1}$ tend to deviate from the baryonic Tully-Fisher relation, a tight scaling relation between the baryonic mass content and flat rotational velocity of a galaxy \citep{2005ApJ...632..859M}, as shown in \citet{2019ApJ...884L..11O,2012ApJ...755..107D,2022MNRAS.517...99R}. For example, the massive spirals like NGC~1961 and NGC~6753 (stellar mass $\rm \gtsim10^{11}M_{\odot}$) with moderate star formation of $\rm \sim 15.5 M_{\odot}yr^{-1}$ and $\rm \sim 11.8 M_{\odot}yr^{-1}$ respectively are reported to contain $\rm \sim 30-50$ per cent fewer baryons in their halos than what is expected from the cosmic baryon fraction \citep{2013ApJ...772...97B}. Similarly, another massive disc galaxy UGC~12591 contains $\rm \sim85$ per cent fewer baryons than that expected from the cosmic mean having a star formation rate of $\rm \sim0.638 M_{\odot}yr^{-1}$ \citep{2022MNRAS.517...99R}. This deficiency in baryons in these galaxies also contributes to their deviations from the baryonic Tully-Fisher relation \citep{2022MNRAS.517...99R}, emphasising their inability to condense halo baryons into stars. So, knowing the different assembly histories of such galaxies is extremely important, and to be able to constrain the star formation over the period, starting from the early Universe, of such galaxies would give us a significant understanding of their evolution.

%Multi-wavelength data on these massive galaxies is necessary for understanding the complex and wide-varying evolution trajectory of these massive objects. 

In this paper, we report the ultraviolet observations of five of the most massive spiral galaxies known, in combination with data at other wavelengths, to address some of the most critical questions in this field, like how the cooling of the circum-galactic gas effect star formation in such galaxies,  how star formation is fuelled or quenched, and what are the dominant feedback mechanisms for quenching the star formation. %- morphological versus AGN feedback?% 
These galaxies are extremely massive with stellar masses $\rm > 10^{11}M_{\odot}$ and they also are fast rotators with rotational velocities $\rm >300 km~s^{-1}$. Here, we investigate the star formation history, nature of the star-forming regions, signatures  and  possible AGN/radiative feedback effects on the star formation, etc, accounting for the total baryon budget in these massive, fast-rotating spiral galaxies in the local Universe. 

The structure of this paper is as follows: In \S2, we describe the observations and data. The data reduction strategy for the UVIT observations, along with the hierarchical structuring of star-forming regions and Spectral Energy Distribution (SED) fitting processes taking FUV to FIR observations of the objects into account, are described in \S3. In \S4, we present our discussion, and the study's conclusions are outlined in \S5. Note that, in our discussion, we explore our objects through the parameters found from direct observations, scaling relations and also from fitting appropriate models to observable data.  In our calculations we use the following cosmological parameters; $\rm H_{0} = 69.6\,km\,s^{-1}Mpc^{-1}$, $\rm \Omega_{M} = 0.286$ and $\rm \Omega_{vac} = 0.714$.
%UVIT consists of two identical telescopes consisting of FUV, NUV, and VIS bands aligned to observe the sky in a field of view of $\sim 28$ arcmin simultaneously. Out of two telescopes, one only has the FUV band while the other has NUV and VIS bands combined and the optics work with the help of a beam-splitter. 

%%%%%%%%%%%%%%%%%

\begin{table*}
    \centering
    \begin{tabular}{ccccc}
    \hline\\[-1.5ex]
         Object & Band & Wavelength & Exposure time & Observation Id \\
          &  & $(\lambda)~\text{\AA}$ & (sec) & \\
         \hline
         NGC~1030 & F154W & 1541 & 1911.445  &A05$\_$225T04$\_$9000002400 \\%& 02-Oct-2018 \\   
         NGC~1961 & F154W & 1541 & 1900.023  &A05$\_$225T02$\_$9000002510 \\%& 13-Nov-2018 \\
         NGC~4501 & F154W & 1541 & 1608.817  &A04$\_$167T05$\_$9000002052 \\%& 25-Apr-2018 \\
         NGC~5635 & F154W & 1541 & 8068.550 &A04$\_$167T02$\_$9000001974 \\%& 13-Mar-2018 \\
         NGC~5635 & N242W & 2418 & 8191.425 &A04$\_$167T02$\_$9000001974 \\%& 13-Mar-2018 \\
         NGC~266 & F154W & 1541 & 7449.227 &A04$\_$167T04$\_$9000001734 \\%&29-Nov-2017\\
         NGC~266 & N242W & 2418 & 7493.291 &A04$\_$167T04$\_$9000001734 \\%& 29-Nov-2017\\
         \hline
    \end{tabular}
    \caption{The details of observations for the five massive spiral galaxies are mentioned here.}
    \label{tab:data}
\end{table*} 

\begin{figure*}
    \centering
    \includegraphics[width=1.0\textwidth]{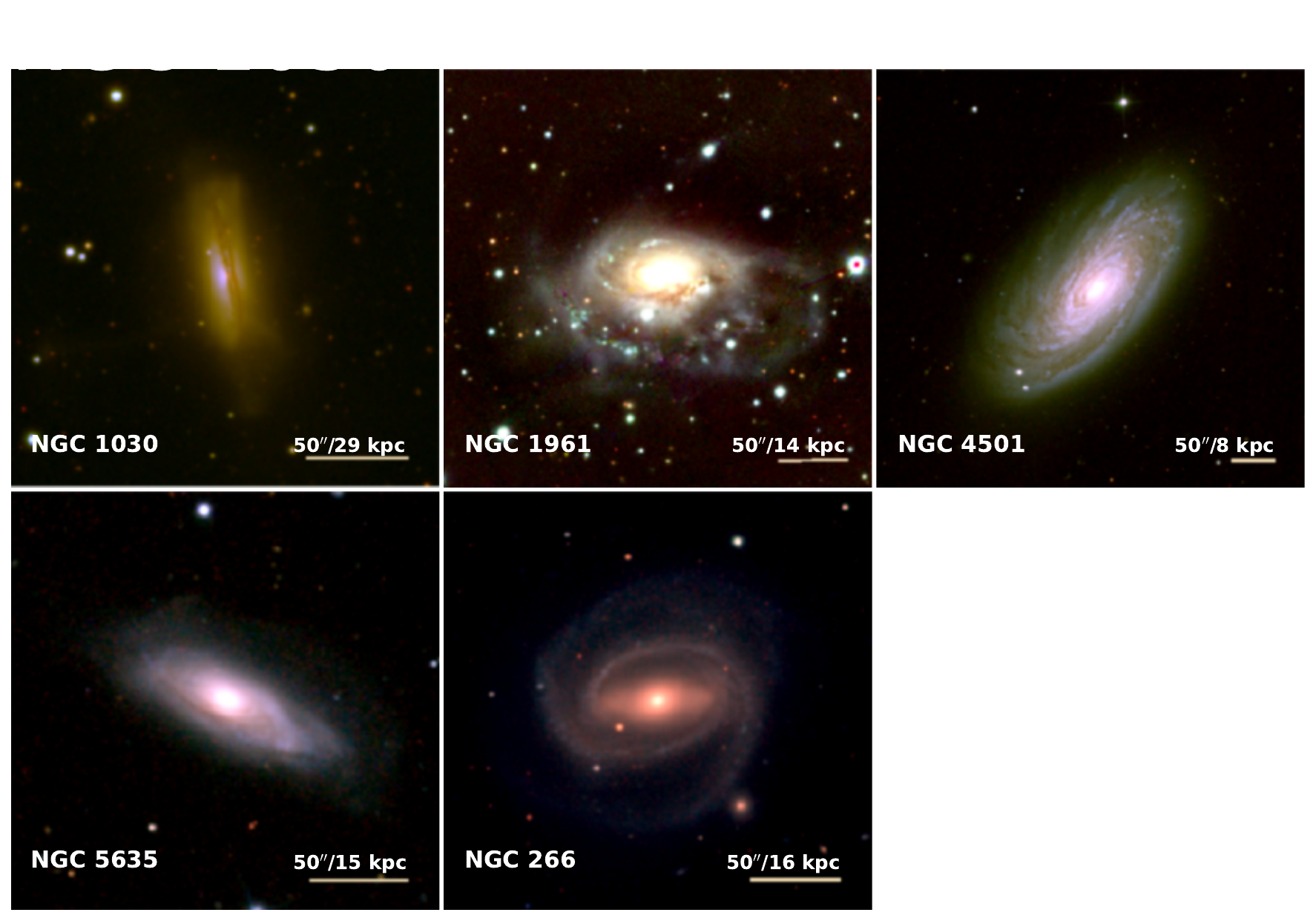}
    \caption{In the above figure, we show the RGB images of the five massive late-type spirals (Table~\ref{tab:data}). The images are made using optical data ($\sim 0.4-0.9 \micron$) in the following configurations: (a) NGC~1030: DECaLS z+r+g, (b) NGC~1961: PanSTARRS1 z+r+g, (c) NGC~4501: SDSS z+r+g, (d) NGC~5635: SDSS z+r+g and (e) NGC~266: SDSS z+r+g. Note that, the scale shown in the lower-right of each image is 50 arcsec.}
    \label{fig:optical_images}
\end{figure*}
\begin{figure*}
    \centering
    \includegraphics[width=1.05\textwidth]{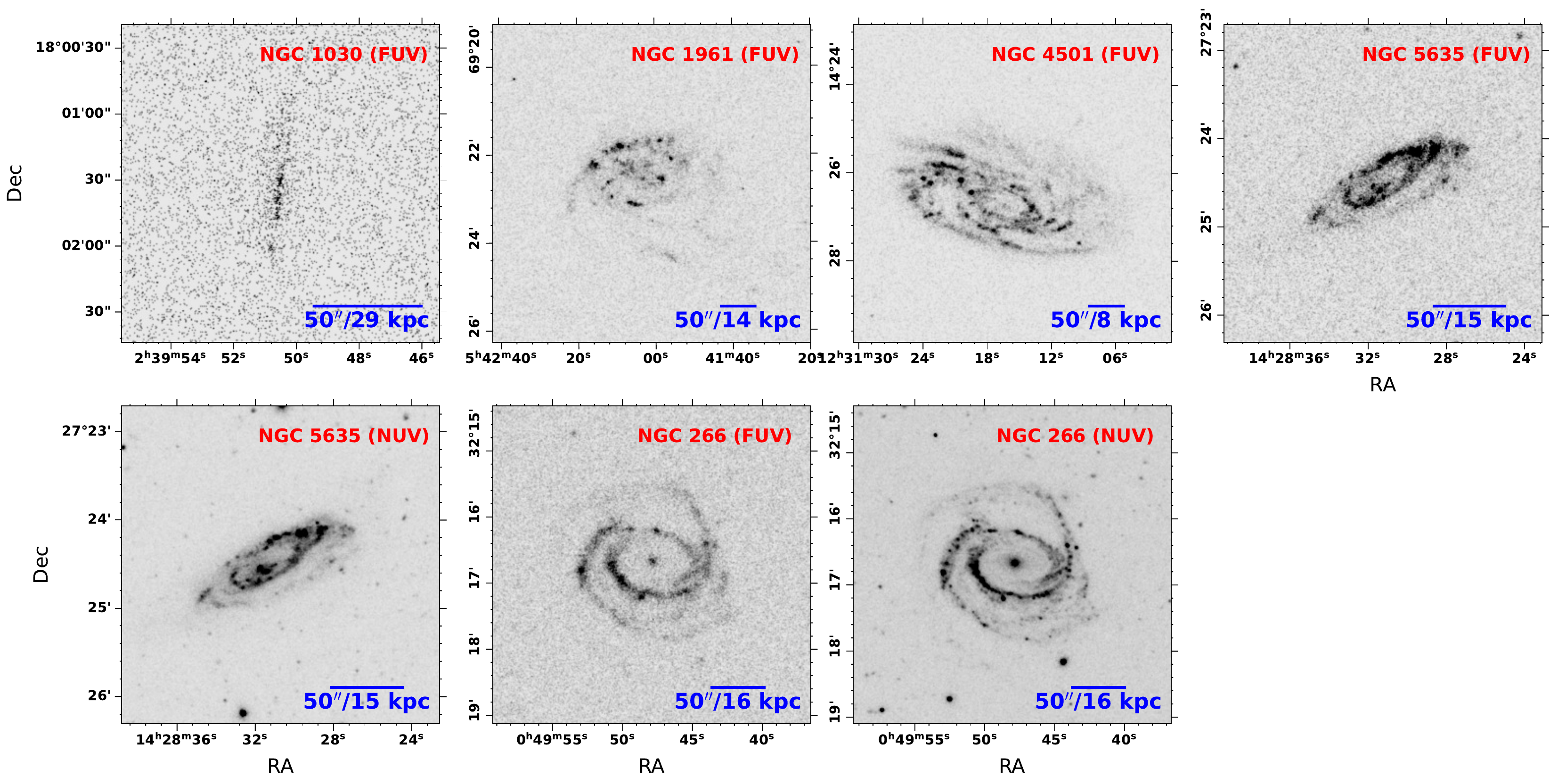}
    \caption{Here, in the above figure, we show the calibrated images of our objects from UVIT Level-1 data processed in CCDLAB \citep{2017PASP..129k5002P}. We have FUV F154W ($\sim 0.1541\micron$) observations for all five spirals and NUV N242W ($\sim 0.2420\micron$) observations for the objects NGC~5635 and NGC~266. In FUV band, the integration times of the objects are $\rm T_{int}\sim$ 1.9, 1.9, 1.6, 8.1 and 7.4 ks for NGC~1030, NGC~1961, NGC~4501, NGC~5635 and NGC~266 respectively. In NUV band, $\rm T_{int}\sim$ 8.2, 7.5 ks for NGC~5635 and NGC~266 respectively. Note that, the scale shown in the lower-right of each image is 50 arcsec.}
    \label{fig:data}
\end{figure*}
%%%%%%%%%%%%%%%%%

%In this paper, we analyse the ultraviolet data of our five objects observed by the UltraViolet Imaging Telescope (UVIT) to study the star formation history of these objects. We use the far-ultraviolet F154W band %(Table~\ref{tab:data}, Figure~\ref{fig:data}) %
%images to study the young star-forming regions in these objects considering the star-forming clump structures to be hierarchical in nature with top-to-down approach identifying brightest star-forming clumps and then adding fainter clumps in the subsequent steps. %(Figure~\ref{fig:clump_structures}, Figure~\ref{fig:dendrogram_ngc5635_fuv}).% 
%In addition to that, we use UVIT far-ultraviolet (F154W) and near-ultraviolet (N242W) data along with other archival data of GALEX, SDSS, WISE etc. (see Section~\ref{sec:observations}) of our objects covering wide wavelength from far-ultraviolet to far-infrared (FIR) to find the spectral energy distribution of the objects to estimate the present-day instantaneous star formation rates, the peak age of star formation history of the objects, stellar mass etc. (Figure~\ref{fig:sed_fitting}, Table~\ref{tab:sed_table}).%

\section{Data}\label{sec:data}
%%%%%%%%%%%%%%%%%%%%%%%
\begin{table*}
    \centering
    \begin{tabular}{lllllll}
    \hline \\[-1.5ex]
        Parameters & Units & NGC~1030 & NGC~1961 & NGC~4501 & NGC~5635 & NGC~266 \\
        \hline \\[-1.5ex]
         R.A.&$-$&02h 39m 50.600s&05h 42m 4.633s&12h 31m 59.153s&14h 28m 31.762s&00h 49m 47.815s \\
         Dec.&$-$&+18d 01m 27.40s&+69d 22m 42.41s&+14d 25m 13.15s&+27d 24m 32.23s&+32d 16m 39.79s \\
         T&$-$&Sc&SABb&Sb&Sb&Sab (Barred) \\
         $\rm \it{z}$&$-$&0.02851&0.01312&0.00762&0.01440&0.01550 \\
         $\rm v_{rot}$&$\rm km~s^{-1}$&371$\rm^{a}$&402$\rm^{b}$&320$\rm^{c}$&386$\rm^{a}$&363$\rm^{d}$ \\
         $\rm \sigma$&$\rm km~s^{-1}$&$-$&$242.8\pm11.5$&$166.2\pm7.1$&$-$&$229.1\pm7.2$ \\
         $\rm \beta$&$-$&$-$&S3&S2&S3&S3b\\[3pt]
         \hline\\[-1ex]
%        Pixel scale&&&&& \\
%        Aperture radius&&&&& \\
%         $\rm A_{FUV}$&&&&&\\
%         $\rm A_{NUV}$&&&&&\\
         $\rm m_{FUV}$&mag&$17.66\pm0.01$&$14.33\pm0.01$&$13.60\pm0.01$&$16.18\pm0.01$&$15.78\pm0.01$ \\
         $\rm m_{NUV}$&mag&$17.34\pm0.05$&$13.49\pm0.03$&$12.86\pm0.03$&$15.87\pm0.002$&$15.15\pm0.002$ \\
         Aperture &$\rm kpc^{2}$&$1.33\times10^{4}$&$1.01\times10^{4}$&$6.24\times10^{3}$&$7.77\times10^{3}$&$5.04\times10^{3}$\\
         $\rm SFR_{FUV}$&$\rm M_{\odot}yr^{-1}$&$0.82\pm0.01$&$4.00\pm0.05$&$2.40\pm0.02$&$0.80\pm0.01$&$1.36\pm0.02$ \\
         $\rm SFR_{FUV}^{corr}$ &$\rm M_{\odot}yr^{-1}$& $5.15\pm0.87$ & $13.66\pm2.04$ & $13.51\pm2.24$ & $1.43\pm0.13$ & $2.54\pm0.25$ \\
         $\rm \Upsilon^{*}_{K}$&$\rm M_{\odot}/L_{\odot}$&0.86&0.78&0.89&0.99&0.83 \\
         $\rm M_{K}$&$\rm M_{\odot}$&$4.92\times10^{11}$&$4.38\times10^{11}$&$5.80\times10^{11}$&$2.47\times10^{11}$&$4.12\times10^{11}$ \\
         $\rm sSFR_{FUV}^{corr}$&$\rm yr^{-1}$&$(1.05\pm0.18)\times10^{-11}$&$(3.12\pm0.47)\times10^{-11}$&$(2.33\pm0.39)\times10^{-11}$&$(0.58\pm0.05)\times10^{-11}$&$(0.62\pm0.06)\times10^{-11}$\\
         $\rm M_{HI}$&$\rm M_{\odot}$&$1.32\times10^{10}$&$3.58\times10^{10}$&$7.09\times10^{9}$&$1.27\times10^{10}$&$7.99\times10^{9}$ \\
         $\rm M_{H2}$ &$\rm M_{\odot}$&$(9.64\pm1.72)\times10^{9}$&$(7.88\pm2.15)\times10^{10}$&$(1.28\pm0.21)\times10^{10}$&$(2.29\pm0.38)\times10^{10}$&$(1.76\pm0.48)\times10^{10}$ \\
         $\rm M_{gas}$&$\rm M_{\odot}$&$(3.16\pm0.24)\times10^{10}$&$(1.58\pm0.30)\times10^{11}$&$(2.75\pm0.29)\times10^{10}$&$(4.90\pm0.52)\times10^{10}$&$(3.55\pm0.66)\times10^{10}$\\
         %$\rm \Sigma_{SFR}^{FUV}$&$0.038\pm0.008$&$0.098\pm0.034$&$0.066\pm0.024$&$0.011\pm0.004$&$0.014\pm0.003$ \\
         $\rm   M_{bh}$&$\rm M_{\odot}$&$(1.45\pm0.37)\times10^{8}$&$(3.00\pm0.60)\times10^{8}$&$(6.01\pm1.21)\times10^{7}$&$(6.31\pm1.48)\times10^{7}$&$(2.34\pm0.37)\times10^{8}$\\
         $\rm M_{halo}$&$\rm M_{\odot}$&$1.68\times10^{13}$&$2.16\times10^{13}$&$1.09\times10^{13}$&$1.91\times10^{13}$&$1.59\times10^{13}$ \\
         $\rm T_{vir}$&keV&$0.43$&$0.50$&$0.32$&$0.46$&$0.41$ \\
         \hline
    \end{tabular}
    \caption{Here, in the upper panel of the table, we mention some properties related to the galaxies. The mentioned parameters are: R.A. = Right Ascension (J2000), Dec. = Declination (J2000), T = morphology, $\it{z}$ = red-shift, $\rm v_{rot}$ = rotational velocity and, $\rm \sigma$ = central velocity dispersion and, $\rm \beta$ = activity class of AGN as per HyperLeda where S2, S3 and S3b denote Seyfert II, LINER and LINER with broad Balmer lines classes of AGNs respectively. In the lower panel, we mention the derived properties of our objects. The mentioned parameters are: $\rm m_{FUV}$ = FUV AB magnitude (corrected for foreground extinction), $\rm m_{NUV}$ =  NUV AB magnitude (corrected for foreground extinction), A = aperture area, $\rm SFR_{FUV}$ = star formation rate, $\rm SFR_{FUV}^{corr}$ = star formation rate from corrected FUV luminosty, $\rm \Upsilon^{*}_{K}$ = K-band stellar mass-to-light ratio, $\rm M_{K}$ = K-band stellar mass, $\rm sSFR_{FUV}^{corr}$ = $\rm SFR_{FUV}^{corr}/M_{K}$, specific star formation rate,  $\rm M_{HI}$ = atomic hydrogen mass, $\rm M_{halo}$ = virial/halo mass, $\rm T_{vir}$ = virial temperature of the hot gas. The references for the rotational velocities $\rm v_{rot}$ are, a: \citet{1988A&A...203...28S}, b: \citet{1979ApJ...230...35R}, c: \citet{1991A&AS...90..121C}, d: \citet{2005MNRAS.362..127G}.}
    \label{tab:all_info}
\end{table*}
%%%%%%%%%%%%%%%%%%%%%

\subsection{Sample selection} \label{sec:sample_selection}

Massive galaxies with high rotational velocities ($\gtrapprox 300 \kms$) have been under study for their departures from the long-established baryonic Tully-Fisher relation. In order to investigate their baryon-to-star conversion, we have observed five such extremely massive  galaxies in UVIT (Table.~\ref{tab:data}) as a pilot sample with outer Keplerian disc rotation velocities $\rm v_{rot} > 300$ km s$^{-1}$ (\citet{1988A&A...203...28S}; \citet{1979ApJ...230...35R}; \citet{1991A&AS...90..121C}; \citet{2005MNRAS.362..127G}, Table~\ref{tab:all_info}) for this study. These galaxies  have K-band stellar masses $> 10^{11}$ \Msun and virial masses $> 10^{13}$ \Msun (see Table~\ref{tab:all_info}, Section~\ref{sec:baryons} and Section~\ref{sec:sfe}). Previously, we published a detailed analysis of one such  massive galaxy, UGC~12591 \citep{2022MNRAS.517...99R}, which has $\rm v_{rot} \approx 500 \kms$. The target galaxies are late-type spirals in the morphological sequence characterised in HyperLeda \citep{2014A&A...570A..13M}. 
%Moreover, these galaxies that are part of our sample have maximum flat-rotational velocity $\rm v_{rot}\gtrapprox 300 kms^{-1}$. 
To see their morphological appearances and other hidden features, we have made color images of these target galaxies in the optical band, which are shown in Fig.~\ref{fig:optical_images}. These images are made using archival data of z, r and g bands from the Dark Energy Camera Legacy Survey (DECaLS), Panoramic Survey Telescope and Rapid Response System (Pan-STARRS) and Sloan Digital Sky Survey (SDSS). From Fig.~\ref{fig:optical_images}, it is evident that the object NGC~266 has the face-on view with an inclination angle (the angle between the line of sight and the polar axis of the galaxy) of $\rm \sim 16$ degree, and the other objects are almost edge-on with inclination angles of $\sim$ 47, 63, 68 and 73 degrees for NGC~1961, NGC~4501, NGC~1030 and NGC~5635 respectively. Note that the inclination angles for our objects have opted from HyperLeda, where the apparent flattening ($\rm r_{25}$) and morphological type (t) of a galaxy are used to find out the inclination (i) from the following relation,

\begin{equation}
\rm sin^{2}(i) = \frac{1-10^{-2log(r_{25})}}{1-10^{-2log(r_{0})}} 
\end{equation}

where $\rm log(r_{0}) = 0.43 + 0.053t$ for $\rm t=[-5,7]$ and $\rm log(r_{0})=0.38$ for $\rm t>7$ (see HyperLeda for more details; \citealt{2014A&A...570A..13M}). The dust lanes are clearly visible in NGC~1030 due to its edge on view. It has a rectangular shape with growing asymmetries visible at the edges \citep{2004A&A...417..527L}. NGC~1961 is classified as an intermediate spiral galaxy, objects falling between barred and unbarred spiral galaxies in the morphological classification, with a bar not well defined, which is evident from the optical color image. The highly asymmetric spiral arms of this object are signatures of possible strong recent interaction(s) \citep{1983MNRAS.202P..21G}, specifically due to the stripping of gas from the gravitational potential of the galaxy by $\rm \sim 10^{7}K$ hot intergalactic medium as proposed by \citet{1982A&A...115..293S}. Any asymmetry in NGC~4501 is not very dominant in the optical image. However, we can see the bright spiral arms embedded into dusty regions throughout the galaxy. However, \citet{2008A&A...483...89V} show the existence of early-stage ram-pressure stripping in the galaxy. A clear asymmetry can be seen in the light distribution for the object NGC~5635, indicating possible recent interaction also shown by \citet{1988A&A...203...28S}. The tight-arm and barred spiral galaxy NGC~266, which is part of a group consisting of six low mass galaxies \citep{2013ApJ...772...97B}, is also found to be interacting with a Seyfert II galaxy Mrk~348 \citep{2001ASPC..240..657H}.

In Table~\ref{tab:all_info}, we show some important parameters for our sample galaxies. Some are obtained from the literature, shown in the upper panel of the table and these are coordinates of the objects (R.A, Dec.), morphological classification T, redshift $\it{z}$, flat rotational velocity $\rm v_{rot}~(km~s^{-1})$, central velocity dispersion $\rm \sigma~(km~s^{-1})$ and activity class $\rm \beta$ of the active galactic nuclei (AGN) of the objects. In the lower panel, we show a list of parameters derived in this work directly from observations or indirectly using scaling relations. The FUV and NUV magnitudes ($\rm m_{FUV}$ and $\rm m_{NUV}$) of the objects are estimated using UVIT data within the apertures estimated following the process described in Section~\ref{sec:photometry}. The FUV star formation rates $\rm SFR_{FUV}$ are estimated using the corresponding FUV magnitudes/luminosities and the relation provided by \citet{1998ARA&A..36..189K} for Salpeter initial mass function (IMF) \citep{1955ApJ...121..161S} (see Section~\ref{sec:star_formation_hierarchy}). The star formation rates corrected for internal dust attenuation of the galaxies are denoted by $\rm SFR_{FUV}^{corr}$ and are estimated considering the total infrared radiations into account following the process described in Section~\ref{sec:sfr}. The K-band mass to light ratios $\rm \Upsilon_{K}^{*}$ of the objects are calculated using specific colors (Section~\ref{sec:baryons}) and the scaling relations provided by \citet{2003ApJS..149..289B}, and we have used them to calculate the K-band stellar masses $\rm M_{K}$ of our objects from 2MASS K-band luminosities. The specific star formation rates $\rm sSFR_{FUV}^{corr}$ are calculated from the dust-corrected star formation rates and the K-band stellar masses as $\rm SFR_{FUV}^{corr}/M_{K}$. The neutral hydrogen masses $\rm M_{HI}$ are calculated from 21cm integrated line flux following the relation provided in Section~\ref{sec:baryons}. On the other hand, the molecular hydrogen masses $\rm M_{H2}$ are found out using the estimated neutral hydrogen mass multiplied by scale factors provided by \citet{1989ApJ...347L..55Y}. The black hole masses $\rm M_{bh}$ are calculated indirectly from the central velocity dispersions $\rm \sigma$ or the stellar masses of the objects (see Section~\ref{sec:bh_sfr} for the relations used and the references). The halo masses $\rm M_{halo}$ are estimated under the assumption of flat circular velocity upto the virial radius within which the mean density is 200 times that of the critical density of the universe at that redshift (more on this in Section~\ref{sec:sfe}). $\rm T_{vir}$ denotes the temperature of the gas in the hot halo of the galaxies, which is considered to be a virialized system, calculated using the rotational velocity (see Section~\ref{sec:baryons}).

\subsection{Observations}\label{sec:observations}
The data were taken by ASTROSAT UVIT under the proposals $A04\_167$ and $A05\_225$ (Principal Investigator  Joydeep Bagchi). The objects are observed in the photon counting mode, and the details of the observations are tabulated in Table~\ref{tab:data}. The  observations were carried out to study the young star formation in our objects.
%{F154W ($\rm \lambda_{eff}\sim1541\text{\AA}$) far-ultraviolet (FUV) and N242W ($\rm \lambda_{eff}\sim2420\text{\AA}$) near-ultraviolet (NUV) bands}.
%The data were taken by AstroSat-UVIT under the proposals $A04\_167$ and $A05\_225$ with Priciple Investigator (PI) Joydeep Bagchi. 
UVIT is a versatile instrument designed to see the sky in FUV ($130\lessapprox \lambda \lessapprox180$ nm), NUV ($200\lessapprox \lambda \lessapprox300$ nm) and in the Visible band (VIS) ($320\lessapprox \lambda \lessapprox550$ nm). UVIT is capable of making images simultaneously in a field of view of $\sim$ 28 arcmins with a resolution of $<$ 1.8 arcsec FWHM. In the two ultraviolet channels, gratings are provided for low-resolution ($\sim 100$) slitless spectroscopy. The focusing optics is configured as twin R-C telescopes, each with a primary mirror with an effective diameter of $\sim 37.5$ 
centimetre. The  FUV and NUV channels provide
science observations, while the VIS channel mainly enables corrections for telescope drifts. The  NUV and FUV channels provide excellent angular resolutions of $\sim$1.2 and $\sim$1.4 arcseconds, respectively,  thereby much improving on their predecessor
GALEX, which had a spatial resolution of $\sim$5 arcseconds. Even though the pixel scale at
the detector plane is $\sim 0.42$ arcsec/pixel, an onboard algorithm enables incoming photons to be
localized on the detector plane with higher accuracy of ${1/8}^{\rm th}$ of a pixel.
Due to this higher resolution capability, UVIT is a powerful tool to observe the young star formation in galaxies in a  well-resolved manner. 

%The specific details of observations are mentioned in Table~\ref{tab:data}.

\begin{figure*}
    \centering
    \includegraphics[width=\textwidth]{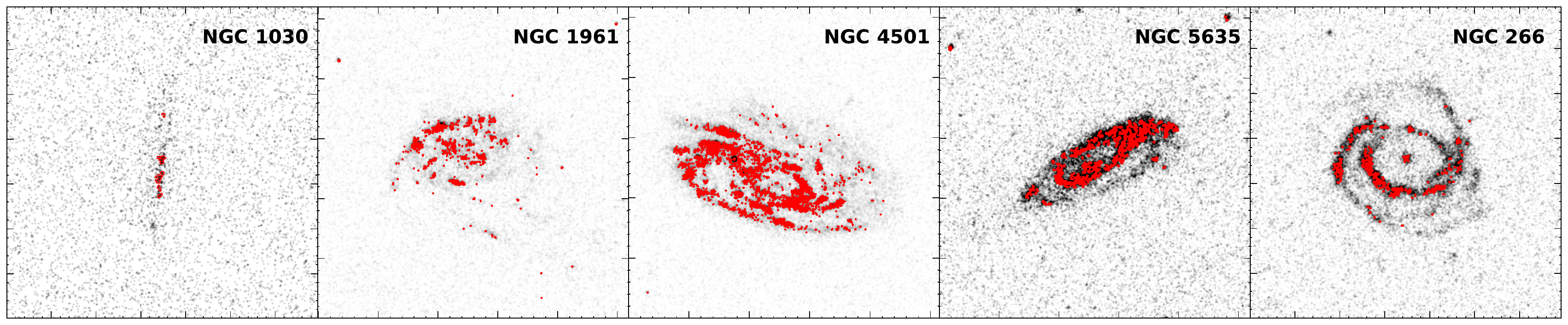}
    \includegraphics[width=\textwidth]{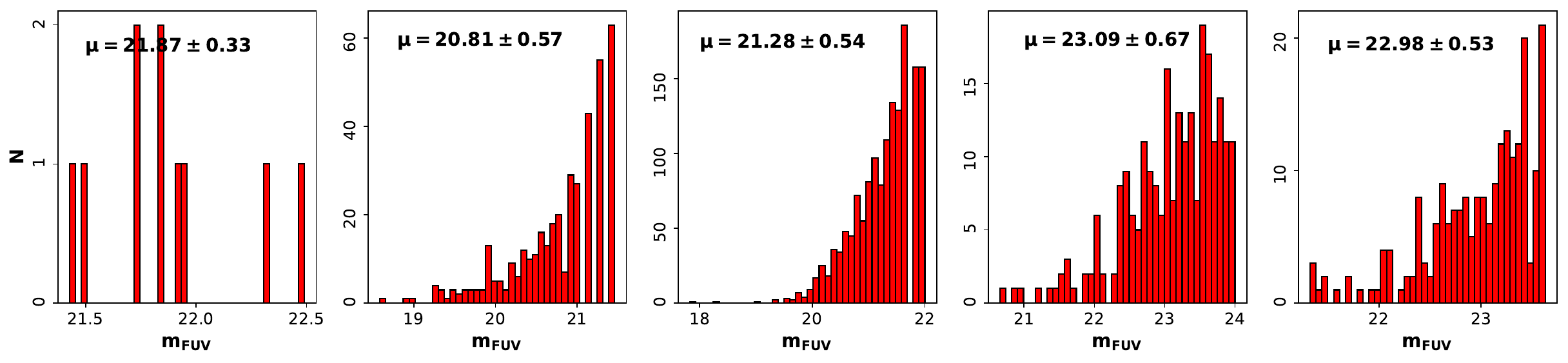}
    \includegraphics[width=\textwidth]{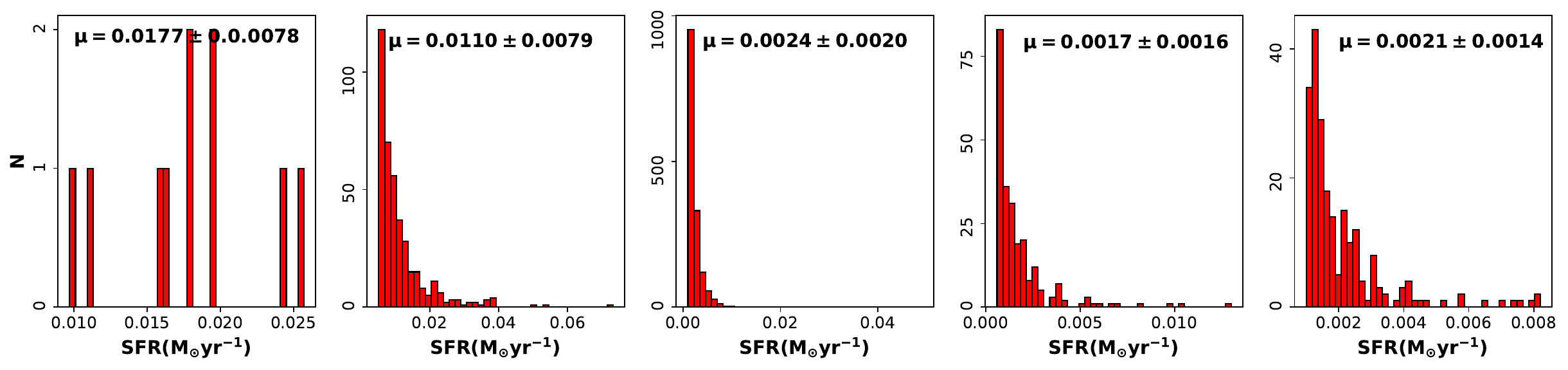}
    \includegraphics[width=\textwidth]{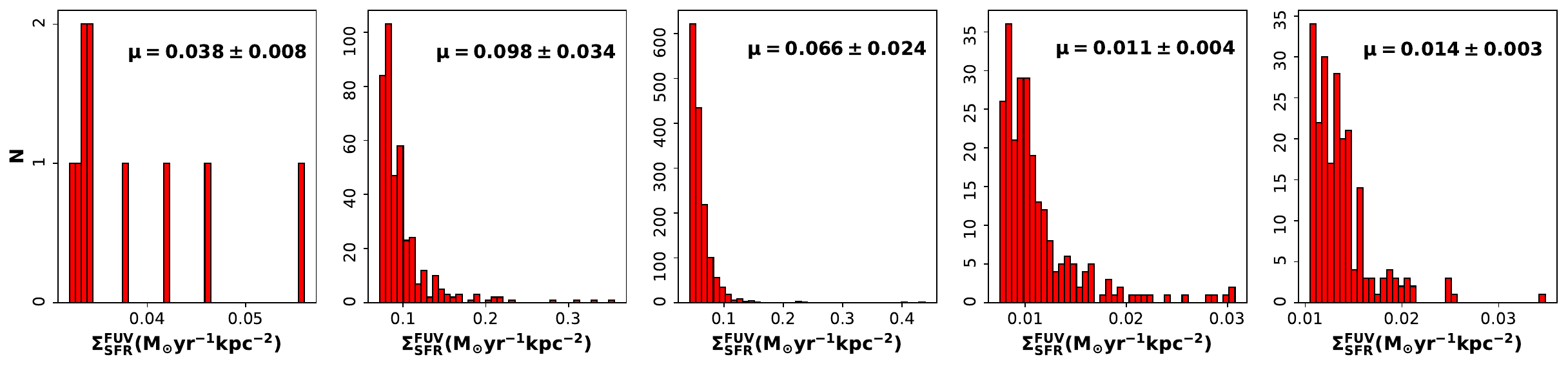}
    \includegraphics[width=\textwidth]{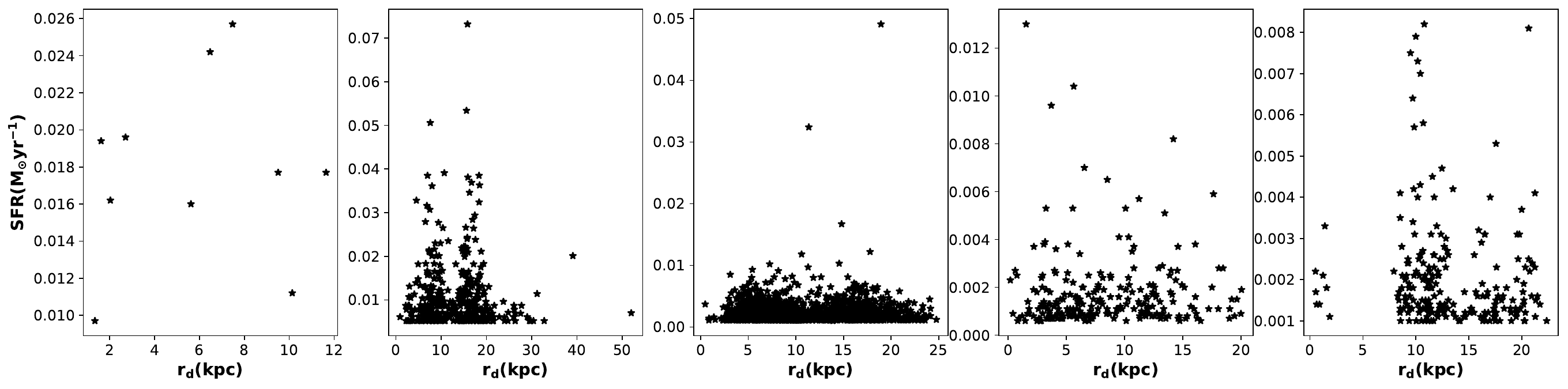}
    \caption{\textit{First row:} Here, we show the young star-forming clumps based on the UVIT far-ultraviolet (FUV) data of the objects that are identified with $ 3\sigma$ detection threshold considering the star-forming clumps to be hierarchical in nature \citep{2019ascl.soft07016R} (Section~\ref{sec:star_formation_hierarchy}). \textit{Second row:} Here, we show the histogram of FUV magnitude ($\rm m_{FUV}$) of the identified clumps in the AB system for the objects with the mean $\rm \mu$ with $\rm 1\sigma$ uncertainties mentioned in the respective plots. \textit{Third row:} Here, we show the histogram of the FUV star formation rate $\rm SFR (M_{\odot}yr^{-1})$ of the identified clumps of the objects with mean $\rm \mu$ and $\rm 1\sigma$ uncertainty.
    \textit{Fourth row:} Here, we show the distribution of FUV star formation rate density $\rm \Sigma_{SFR}^{FUV} (M_{\odot}yr^{-1}kpc^{-2})$ of the objects with the mean $\rm \mu$ and $\rm 1\sigma$ uncertainty. \textit{Fifth row:} Here, the variation of the FUV SFR of the identified clumps with radius $\rm r_{d} (kpc)$ from the centre of the galaxy is shown for each object.}
    \label{fig:clump_structures}
\end{figure*}

%{\bf JB:  WHAT FILTERS WERE USED IN NUV and FUV channels?}
%\vskip 
In this study, in addition to ASTROSAT UVIT data we have also used archival data from the following missions: 
\begin{enumerate}
\item  Galaxy Evolution Explorer (GALEX) (NUV)

\item Sloan Digital Sky Survey (SDSS) (u, g, r, i, z)

\item Dark Energy Camera Legacy Survey (DECaLS) (g, r, i)

\item Panoramic Survey Telescope and Rapid Response System (Pan-STARRS) (g, r, i, z, y)

\item Two Micron All-Sky Survey (2MASS) (J, H, K)

\item  Wide-field Infrared Survey Explorer (WISE) (W1, W2, W3, W4) and

\item Infrared Astronomical Satellite (IRAS) (IRAS1, IRAS2, IRAS3, IRAS4).
\end{enumerate}

\section{Data reduction and analysis}

\subsection{UVIT Level-1 data reduction}\label{sec:data_reduction}

We obtained the Level-1 (L1) data for our objects from the ASTROSAT data archive. We used CCDLAB provided by \citet{2017PASP..129k5002P} to make observational and detector-related corrections to the L1 data to make it useful for scientific purposes. The L1 files for a particular object contain a FITS binary extension involving a table of the centroid list. Each L1 file may contain multiple observational frames of an object that may be unique or repetitive in nature for each filters.

The image processing by CCDLAB is described in brief as follows: (i) In the first step, all the L1 data files are extracted and checked for their scientific usefulness. All of the duplicate and non-useful data are exempted from further analysis. While observing an object, a short observation of the first 20 seconds is done to test the bright object detection. These observations do not contain enough information to be useful scientifically, and these data are disregarded. All of these are done automatically by CCDLAB. The underlying principle is to create unique subsets of data from the extracted L1 files for each filter based on frame number and frame time of the centroid list, (ii) satellite/telescope drift corrections performed by taking VIS channel observations into account, (iii) exposure array correction, and (iv) finding astrometric solution of World Coordinate System (WCS) to assign a coordinate system to the observed objects. Most of the tasks are automated in the pipeline, although manual intervention is needed for some processing parts. For example, the objects having multiple slices of datasets for a particular band needed to be registered manually with point sources for aligning and merging the images so that translational and/or rotational drifts do not affect the quality of the image. The final calibrated images of our objects from UVIT L1 data are shown in Fig.~\ref{fig:data}. In this figure, 
 we show the FUV F154W ($\sim 0.1541\micron$) observations for all target galaxies and NUV N242W ($\sim 0.2418\micron$) observations for the objects NGC~5635 and NGC~266.
 
 In the FUV band, the integration times of the objects NGC~1030, NGC~1961, NGC~4501, NGC~5635 and NGC~266 are $\rm T_{int}\sim$ 1.9, 1.9, 1.6, 8.1 and 7.4 ks, respectively. While, in the NUV band the integration times for objects NGC~5635 and NGC~266 are $\rm T_{int}\sim$ 8.2, 7.5 ks, respectively.

\subsection{Aperture photometry}\label{sec:photometry}

We perform aperture photometry of the objects in each band mentioned in Section~\ref{sec:observations}. The apertures for finding out the flux of the objects is estimated with the IRAF task \textit{ellipse} \citep{1986SPIE..627..733T}. With the UVIT FUV images as the references, the radius of the aperture is defined as the distance at which the intensity of the source matches the background, which is the mean per pixel calculated from regions away from the source. To estimate the background flux per pixel, we use the SAOImageDS9 task \textit{region} \citep{2003ASPC..295..489J}. Depending on the aperture radius, we make image cut-outs for our objects using the IRAF task \textit{imcopy}. The data preferences are as follows: (a) for FUV and NUV, we use UVIT for all objects except NGC~1030, NGC~1961 and NGC~4501, where in the absence of UVIT NUV, we use GALEX NUV data, (b) for optical bands we use SDSS for NGC~4501, NGC~5635 and NGC~266. In the absence of SDSS images, we use Pan-STARRS images for NGC~1961 and DECaLS images for NGC~1030; (c) for near-infrared, we use 2MASS for all objects; (d) for mid-infrared, we use WISE for all objects except NGC~4501 due to absence of image covering our entire aperture, and in place of that we use IRAS1 and IRAS2, (e) for far-infrared we use IRAS3 and IRAS4 for all our objects. Note that, in the absence of the entire image in our defined aperture, we have used python module \textit{reproject} to specifically make flux-conserved image mosaics for 2MASS NGC~5635 and 2MASS NGC~266 collecting image frames containing our objects of interest from the corresponding archive. For the same reason, we also use SDSS mosaics, but in this case, we use the SWARP \citep{2002ASPC..281..228B,2010ascl.soft10068B} script provided by the DR12 Science Archive Server (SAS) to make the mosaics.

To eliminate the contributions from the background sources to the total light within the aperture, we use SExtractor \citep{1996A&AS..117..393B} for the image cut-outs with DETECT$\_$THRESH as 3 and subtract their fluxes from the total flux within the defined aperture. To eliminate the effect of overall background contribution, we calculate the background flux per pixel as the mean flux from regions far from any contaminating sources in the images using SAOImageDS9 as mentioned before. The process is carried out for each data mentioned in Section~\ref{sec:observations}. Note that, to estimate the fluxes, we use relations provided by \citet{2017AJ....154..128T} for UVIT.

Lower wavelength radiations like ultraviolet (UV) light are highly interactive with an interstellar medium consisting of dust grains. The physical dimensions of dust grains are quantified by the power law of \citet*{1977ApJ...217..425M} to be $\rm \sim 0.005-0.25\micron$. Now, with the UV wavelength in the range $\rm \sim 0.01-0.4\micron$ it can be susceptible to absorption or scattering by dust grains if the grain size is greater or comparable to the wavelength, respectively. The observed UV light from the star-forming regions of galaxies can be thought to be reprocessed in two steps: internal extinction and foreground galactic extinction. To account for the galactic extinction, we consider the dust map of \cite{2011ApJ...737..103S} and calculate the band extinctions taking individual reddening and \cite{1989ApJ...345..245C} extinction law into consideration. The extinction in a particular wavelength is given by,

\begin{equation}
 \rm    A_{\lambda} = R_{\lambda} E(B-V)
\end{equation}

Here, $E(B-V)$ is the color excess for the object. The quantity $R_{\lambda}$, the ratio of total to selective total extinction is 3.1 for Milky Way type extinction in V-band \citep{1975A&A....43..133S}.

\subsection{Star formation hierarchy in the galactic discs}
\label{sec:star_formation_hierarchy}

Far-ultraviolet radiations (FUV) from a galaxy are useful tracers of young star-forming populations within the galaxy. FUV can locate the clumps which are hosting star formation in the galaxy over the past $\sim$ 100 Myr. A large star-forming region may contain smaller high-UV-intensity regions with young star formations going on. We consider these star-forming structures to be hierarchical in nature and employ ASTRODENDRO \citep{2019ascl.soft07016R} to identify these regions based on some given constraints. The algorithm of ASTRODENDRO considers an input intensity map (in our case, the UVIT data) to be hierarchical in nature and decompose it like the structure of a tree. The dendrogram is constructed starting with identifying the brightest pixels in the map and progressively adding fainter pixels in the subsequent steps. The dendrogram considered has two types of structures called branches and leaves. A branch can split into another branches and leaves, but leaves do not have any further substructures. All these structures converge into a trunk that has no parent structure (see \citealt{2019ascl.soft07016R} for more details \footnote{https://dendrograms.readthedocs.io/en/stable/}). The large star-forming regions are considered parent structures in ASTRODENDRO, with smaller regions within these as the child structures. We use $\sim(1.8/0.42)$ as the minimum number of pixels based on the resolution and pixel scale of UVIT in order to identify the brightest clumps. The detection threshold for the clumps is set at $ 3\sigma$ level based on the average noise of the data. The young star-forming clumps of the objects identified with 3$\sigma$ detection threshold are shown in Fig.~\ref{fig:clump_structures} (First row). The parameters derived from the catalogue of star-forming clumps for the objects are shown in Table~\ref{tab:astrodendro_stats}, and their histograms are shown in the second, third, and fourth rows in Fig.~\ref{fig:clump_structures}. In the same figure, fifth row, the variation of the FUV SFR of the identified clumps with radius $\rm r_{d} (kpc)$ from the centre of the galaxy is shown for each object.

\begin{figure}
%\begin{minipage}{0.5\textwidth}
\centering
    \includegraphics[width=\columnwidth]{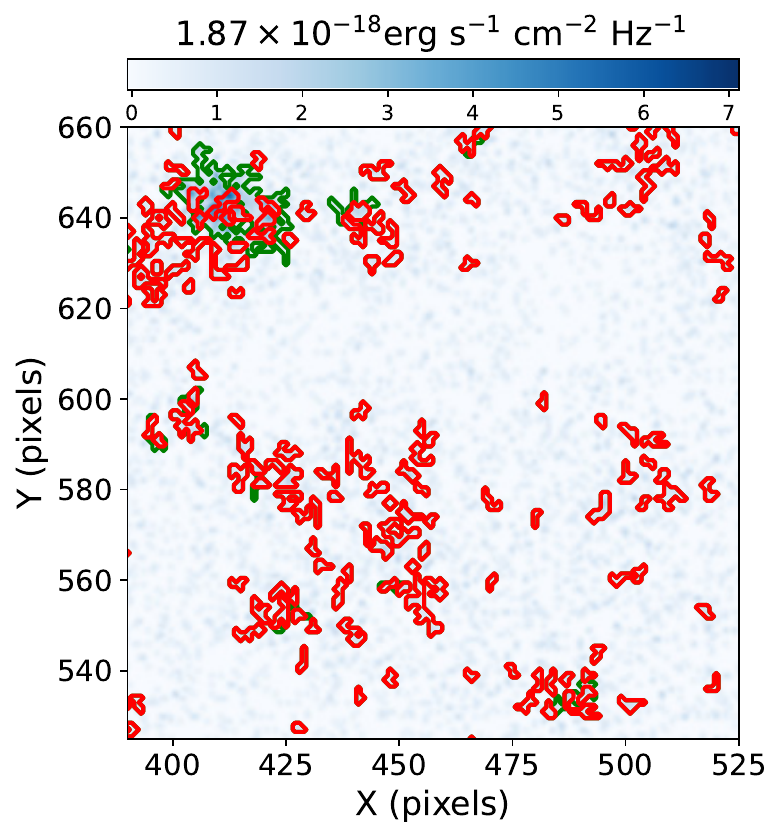}
    \caption{Here, we show a cut-out region ($\rm \sim 57\times57$ arcsec$^{2}$) of star forming clumps (Fig.~\ref{fig:clump_structures}) of the object NGC~1961. The green contours indicate the branches and the red contours indicate the child structures or the leaves. It can be seen that many of the leaves are isolated structures showing isolated regions of young star formation. Note that the image pixel values are in the units of $\rm 1.87\times10^{-18}erg~s^{-1}~cm^{-1}~Hz^{-1}$.}
    \label{fig:ngc1961_cutout_clumps}
\end{figure}

A cut-out image of a region ($\rm \sim 57\times57$ arcsec$^{2}$) of the star-forming clumps in NGC~1961 is shown in Fig.~\ref{fig:ngc1961_cutout_clumps}. The green contours indicate the branches and the red contours indicate the child structures or the leaves. It can be seen that many of the leaves are isolated structures showing isolated regions of young star formation. Note that the image pixel values are in the units of $\rm 1.87\times10^{-18}erg~s^{-1}~cm^{-1}~Hz^{-1}$. The few parameters of the output catalogue of ASTRODENDRO that were made use of are the position coordinates of the clumps, the exact area of the child structures, the flux enclosed by that area, and the effective radius of the clumps. Note that, we have considered the child structures for estimating our parameters as they are more resolved than the parent structures based on the resolution of UVIT.

%The dendrogram of the parent and child structures for a region of NGC~5635 are shown in Fig.~\ref{fig:dendrogram_ngc5635_fuv}.

 %%%%%%%%%%%%%%%%%%%%%%%%%%%%%%%%%

%\end{minipage}
%\begin{minipage}{0.5\textwidth}
%\centering
%    \includegraphics[width=\textwidth]{dendrogram_structure_flux_ngc5635_fuv.eps}
%    \caption{\ray{Have to remove it.} Here, we show how the parent and child structures are distributed in the hierarchical star formation algorithm of ASTRODENDRO \citep{2019ascl.soft07016R}. We show a distribution of 40 structures out of 238 child structures identified in NGC~5635 (see Table~\ref{tab:astrodendro_stats}). The upper structures with the highest fluxes are called the child structures or the leaves (shown in the left graph), and all leaves merge into a few parent structures (can be seen from the right graph), also known as trunks that have no further parent structures. The flux mentioned here is in the units of $\rm 4.40\times10^{-19} erg~s^{-1}~cm^{-2}~Hz^{-1}$.}
%    \label{fig:dendrogram_ngc5635_fuv}
%\end{minipage}
%\end{figure}

%%%%%%%%%%%%%%%%%%%%%%%%%%%%%%%%%

\begin{figure*}
    \centering
    \includegraphics[width=\textwidth]{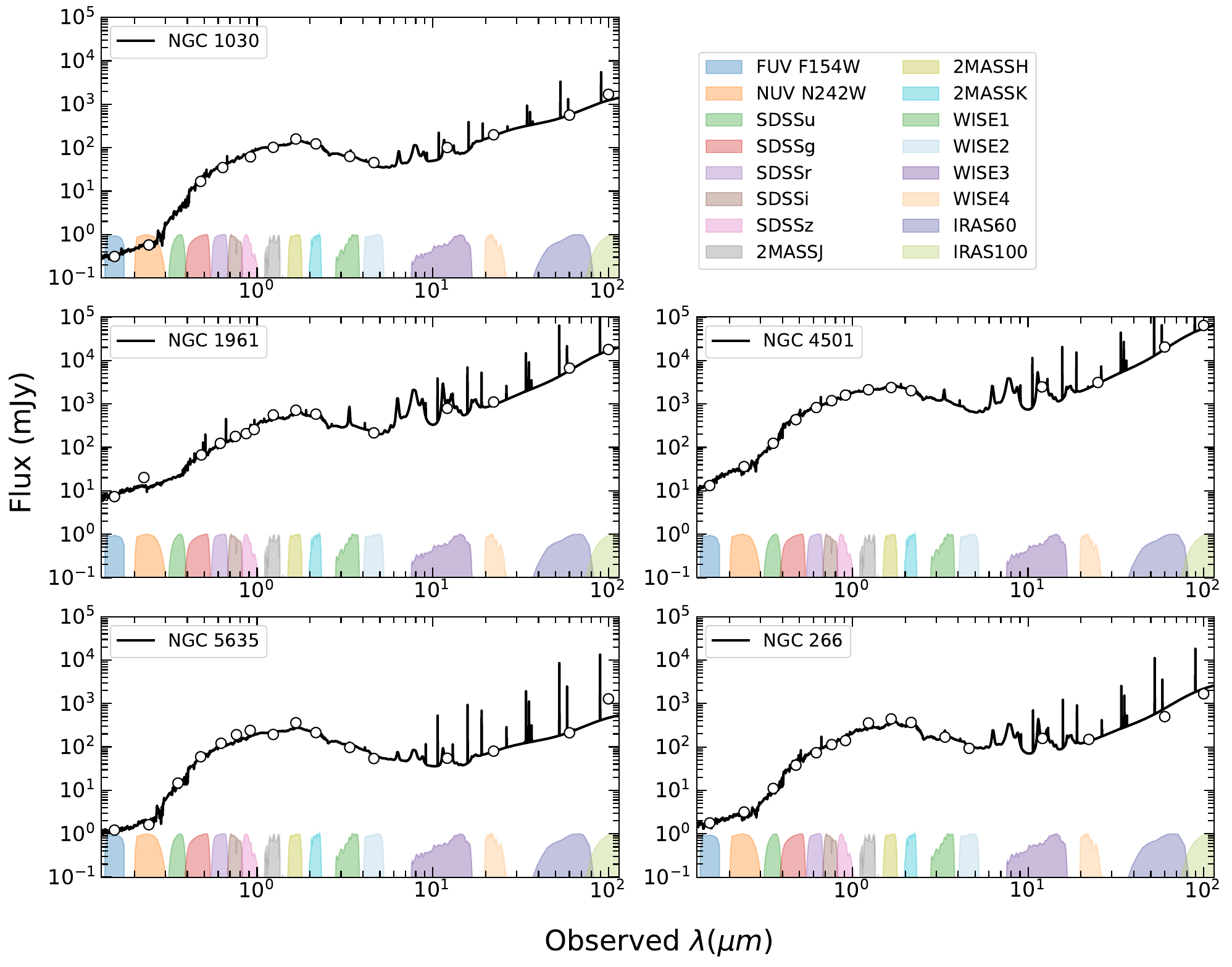}
    \caption{The best-fit models of the spectral energy distributions of the objects using CIGALE are shown. The data points are shown with empty black circles, and the models are shown with black lines. We have employed the following modules to estimate the best-fit results from the FUV-FIR SED: \textit{sfhdelayed} for star formation history SFH, \textit{bc03} for a single stellar population SSP, \textit{nebular} for nebular emission lines, \textit{dustatt$\_$calzleit} for dust attenuation, \textit{dl2014} for dust emission, \textit{fritz2006} for AGN contribution. In the lower part of each graph, we show the bands that have been used to find the SEDs of the objects. Filters from left to right: F154W, N242W, SDSS u, SDSS g, SDSS r, SDSS i, SDSS z, 2MASS J, 2MASS H, 2MASS K, WISE1, WISE2, WISE3, WISE4, IRAS60, IRAS100. The filters with their corresponding colors are also shown in the top right corner of the figure. Note that we have also made use of DECaLS (g, r, z) and Pan-STARRS1 (g, r, i, z, y) images for NGC~1030 and NGC~1961 respectively (Section~\ref{sec:observations}).}
    \label{fig:sed_fitting}
\end{figure*}

 %%%%%%%%%%%%%%%%%%%%%%%%%%%%%%%%%

 %%%%%%%%%%%%%%%%%%%%%%%%%%%%%%%%%
\begin{figure*}
    \centering
    \includegraphics[width=0.49\textwidth]{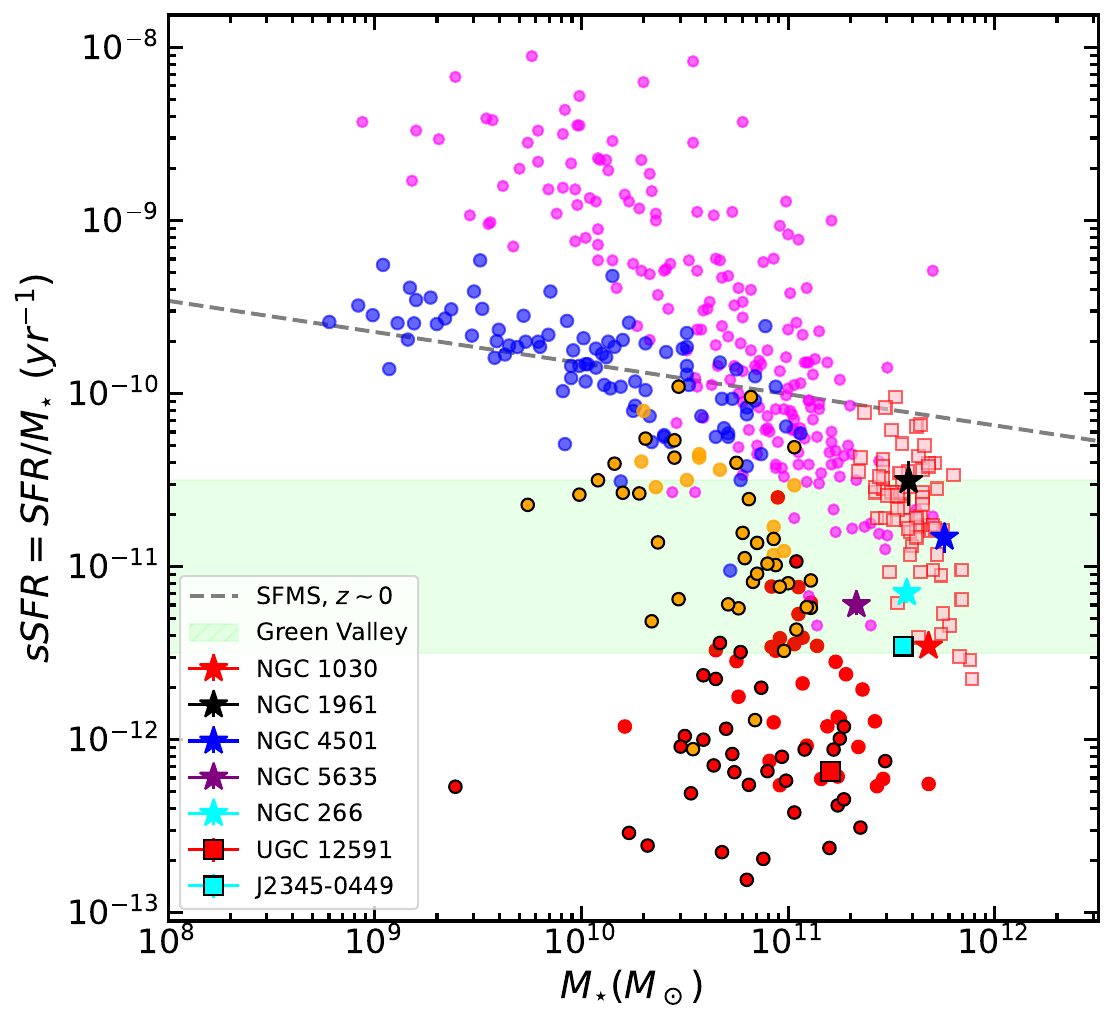}
    \includegraphics[width=0.5\textwidth]{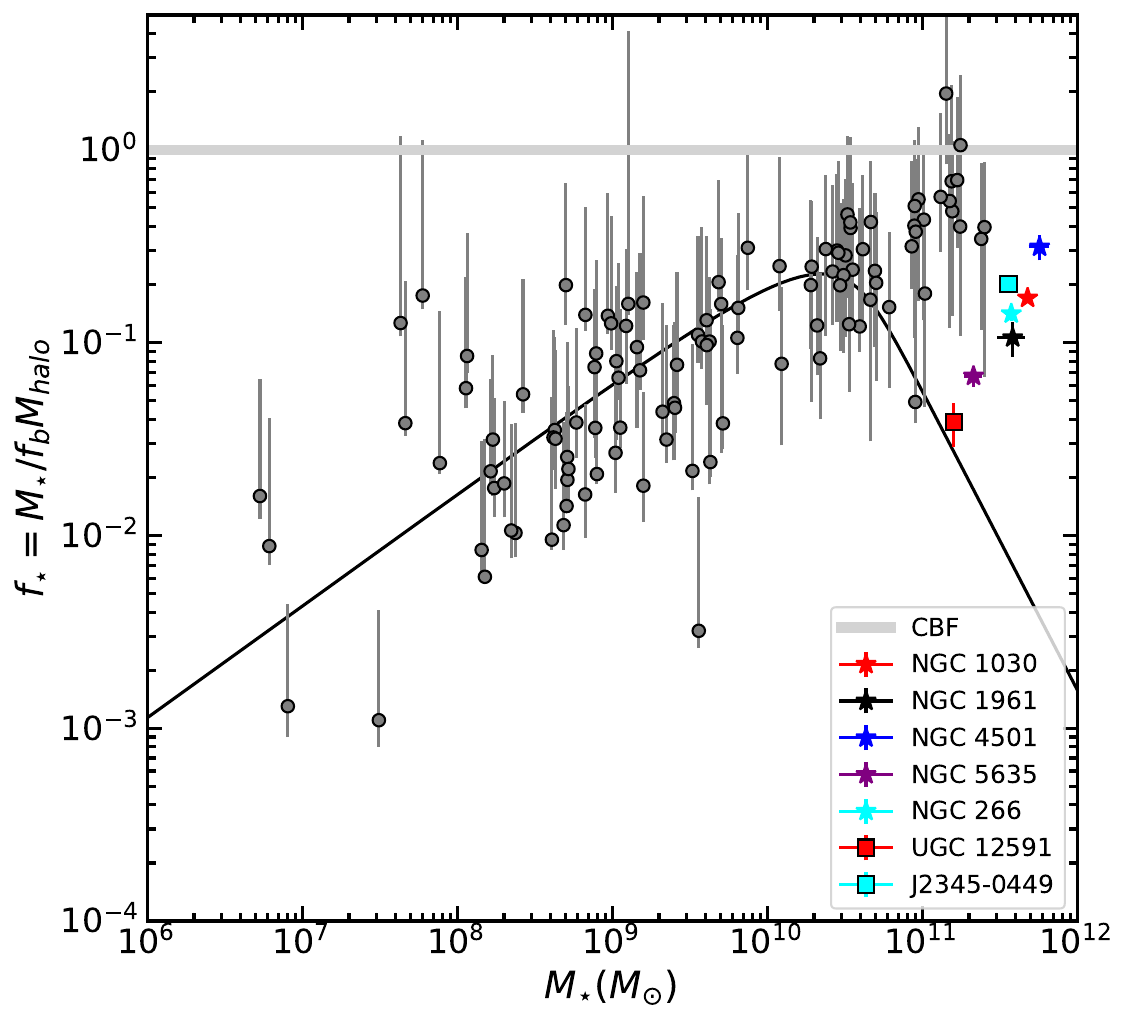}
    \caption{\textit{Left panel:} Variation of specific star formation rate (sSFR in ${\rm yr^{-1}}$) versus the stellar mass ($\rm M_{\star}$ in \Msun) is  shown for galaxies of various types. 
         The dotted line indicates the star-forming the main sequence from $z\sim0$ \citep{2007A&A...468...33E}. Highly Star-forming ultraviolet luminous galaxies are taken from \citet{2007ApJS..173..441H} (filled magenta circles). Super spirals, lenticulars and post-merger galaxies from \citet{Ogle_2019} (orange boxes with red edges). A sample of galaxies with different quenching stages like the star-forming (filled blue circles), mixed (filled orange circles with black edges), nearly retired (filled red circles), fully retired (filled red circles with black edges) and quiescent-nuclear-ring (filled orange circles) taken from \citet{2021A&A...648A..64K} are also shown. The positions of five target galaxies from the present study are shown with star symbols, with instantaneous SFR and stellar mass from SED using CIGALE. The light green shadow indicates the green valley region with $\rm 10^{-11.5}<sSFR (yr^{-1})<10^{-10.5}$ \citep{2016ApJS..227....2S}.
         %One can see that in spite of their high stellar masses, their sSFR is anomalously low, much below the main sequence, and close to the region of most massive  super spirals of \citet{Ogle_2019}, but  above the  highly quenched, fully retired galaxies. 
         \textit{Right panel} shows the variation of the star formation efficiency $f_{\star}$ of the galaxies as a function of its stellar mass $M_{\star}$. The black curve is obtained from the relation between halo mass and the corresponding stellar mass provided by \citet{2013MNRAS.428.3121M} using a multi-epoch abundance matching model. Here, we show the model for redshift $\it{z}\sim 0$. The SPARC (Spitzer Photometry and Accurate Rotation Curves) sample of disc galaxies \citep{2016AJ....152..157L,2019A&A...626A..56P} is shown with black+grey points. In the case of $\rm M_{\star}/M_{halo} = f_{b}$, cosmic baryon fraction (CBF), the star formation efficiency $f_{\star} = 1$ and is shown with the light grey line.}
        \label{fig:ssfr_mstar}
\end{figure*}
%%%%%%%%%%%%%%%%%%%%%%%%%%%%%%%%%

%%%%%%%%%%%%%%%%%%%%%%%%%%%%%%%%%
\begingroup
\setlength{\tabcolsep}{3.8pt}
\begin{table}
    \centering
    \begin{tabular}{llllll} 
    \hline \\[-1.5ex]
    Parameters&NGC~1030&NGC~1961&NGC~4501&NGC~5635&NGC~266\\
    \hline \\[-1.5ex]
    N&10&393&1511&238&219\\
    $\rm m_{FUV}$&21.87&20.81&21.28&23.09&22.98\\
    &$\pm$0.33&$\pm$0.57&$\pm$0.54&$\pm$0.67&$\pm$0.53\\
    SFR&0.0177&0.0110&0.0024&0.0017&0.0021\\
    &$\pm$0.0078&$\pm$0.0079&$\pm$0.0020&$\pm$0.0016&$\pm$0.0014\\
    $\rm \Sigma_{SFR}^{FUV}$&0.038&0.098&0.066&0.011&0.014\\
    &$\pm$0.008&$\pm$0.034&$\pm$0.024&$\pm$0.004&$\pm$0.003\\
    $\rm r_{eff}$&160-317&57-251&34-168&69-310&72-270\\
    \hline
    \end{tabular}
    \caption{In the above table, we mention the mean ($\rm \mu$) and $\rm 1\sigma$ uncertainty of the parameters derived from the catalogue of star-forming clumps of the objects identified with $\rm 3\sigma$ detection threshold in ASTRODENDRO \citep{2019ascl.soft07016R}. The parameters are described below N = the number of clumps, $\rm m_{FUV}$ = magnitude of the identified clumps in the AB system, SFR = star formation rate of the identified clumps in $\rm M_{\odot}yr^{-1}$, $\rm \Sigma_{SFR}^{FUV}$ = star formation rate density of the clumps in $\rm M_{\odot}yr^{-1}kpc^{-1}$, $\rm r_{eff}$ = effective radius (min value-max value) of the star-forming clumps in pc. All the parameters are derived for the area within the apertures mentioned in Table~\ref{tab:all_info}. Note that $\rm m_{FUV}$, SFR and $\rm \Sigma_{SFR}^{FUV}$ shown here are corrected for foreground and internal dust attenuation.}
    \label{tab:astrodendro_stats}
\end{table}
\endgroup

%%%%%%%%%%%%%%%%%%%%%%%%%%%%%%%%%

\begingroup
\setlength{\tabcolsep}{3.0pt}
\begin{table*}
    \centering
    \begin{tabular}{llllllll}
    \hline\\[-1.5ex]
    Modules & Parameters & Units & NGC~1030 & NGC~1961 & NGC~4501 & NGC~5635 & NGC~266 \\
    \hline \\[-1.5ex]
    Star formation history:&(a) $\rm \tau_{main}$&Myr&1$-$8000;20&500$-$5000;5&1$-$5000;20&1$-$8000;20&1000$-$13000;13 \\
     (\textbf{\textit{sfhdelayed}})&(b) Age &Myr&1000$-$13000;13&1000$-$13000;13&500$-$13000;8&5000$-$13000;20&5000$-$13000;8\\
     &(c) $\rm \tau_{burst}$&Myr&[5,10,25,&[1,5,10,&[5, 10, 25,&[1, 5, 10,&[5, 10, 25,\\
     &&&50,100]&25,50,100]&50, 100]&25, 50, 100]&50, 100]\\
     &(d) $\rm Age_{burst}$&Myr&[5, 10, 25,&[5,10,25&[5, 10, 25,&[50, 100, 200,&[5, 10, 25,\\
     &&& 50, 100, 200,&50,100,200&50, 100, 200,&350, 500, 750,&50, 100, 200,\\
     &&&350, 500, 750,&350,500,750&350, 500, 750,&1000,1250, 1500]&350, 500, 750,\\
     &&&1000]&1000]&1000]&&1000]\\
     &(e) $\rm f_{burst}$&$-$&[0, 0.0001, 0.0005,&[0, 0.0001, 0.0005,&[0, 0.0001, 0.0005,&[0.0001, 0.0005, 0.001,&[0, 0.0001, 0.0005,\\
     &&&0.001, 0.005, 0.01,&0.001, 0.005, 0.01,&0.001, 0.005, 0.01,&0.005, 0.01, 0.05,&0.001, 0.005, 0.01,\\
     &&& 0.05, 0.1, 0.25]&0.05]&0.05, 0.1, 0.25]&0.1, 0.25]&0.05, 0.1, 0.25]\\
    Stellar population: &(f) IMF&$-$&Salpeter&Salpeter&Salpeter&Salpeter&Salpeter \\
    (\textbf{\textit{bc03}})&(g) Metallicity &$-$&0.02, 0.05&0.02, 0.05&0.02, 0.05&0.02, 0.05&0.02, 0.05 \\[4pt]
    Dust attenuation: &(h) $\rm E(B-V)_{\rm young}$&$-$&0.1$-$1;5&0$-$3;15&0$-$3;15&0$-$3;15&0$-$1;10 \\
    (\textbf{\textit{dustatt$\_$calzleit}})&(i) $\rm E(B-V)_{\rm old-factor}$ &$-$&0.5$-$1;4&0$-$1;10&0$-$1;10&0$-$1;10&0.1$-$1;10 \\[4pt]
    Dust emission: & (j) $\rm q_{\rm PAH}$&$-$&2.5&6.63&3.90&3.90&7.32\\
%    &&&&&&&\\
    (\textbf{\textit{dl2014}})&(k) $\rm \alpha$ &$-$&2.0&2.2&2.2&2.0&2.2 \\
%    &&&&&&&\\
%    &&&&&&&\\
%    &&&&&&\\
    \hline
    \end{tabular}
    \caption{Here, we mention the grid used in CIGALE to find out the best-fit parameters. The parameters are as follows: (a) $\tau_{main} =$ E-folding time of the main stellar population model, (b) Age = Age of the main stellar population in the galaxy, (c) $\rm \tau_{burst}$ = E-folding time of the late starburst population model, (d) $\rm Age_{burst}$ = Age of the late burst, (e) $\rm f_{burst}$ = Mass fraction of the late burst population, (f) IMF = Initial Mass Function (Salpeter; \citealt{1955ApJ...121..161S}), (g) Metallicity, (h) $E(B-V)_{\rm young} =$ Colour excess of the stellar continuum light for the young population, (i) $E(B-V)_{\rm old-factor} =$ Reduction factor for the E(B-V) of the old population compared to the young one, (j) $q_{\rm PAH} =$ Mass fraction of PAH, (k) $\alpha =$ Power-law slope for $\rm \frac{dU}{dM}\propto U^{\alpha}$. Note that $\rm \tau_{main} = 1-8000;20$ means that $\rm \tau_{main}$ is varied between 1 and 8000 with 20 evenly spaced values as inputs. And also, the parameters for which one single value (e.g. $\rm q_{PAH}, \alpha$) is mentioned, that is, the best fit value for that parameter obtained through multiple iterations.}
    \label{tab:cigale_grid}
\end{table*}
\endgroup
%%%%%%%%%%%%%%%%%%%%%%%%%%%%%%%%%
\begin{table*}
    \centering
    \begin{tabular}{lllllll}
    \hline \\[-1.5ex]
    Parameters & Units & NGC~1030 & NGC~1961 & NGC~4501 & NGC~5635 & NGC~266\\
    \hline \\[-1.5ex]
%    $\it{z}$&&&0.00762&0.0144&0.0155\\
    $\rm SFR_{inst}$& $\rm M_\odot yr^{-1}$&$1.66\pm0.08$&12.04$\pm$2.48&$8.45\pm0.97$&$1.29\pm0.06$&$2.65\pm0.27$\\

%    $\rm SFR_{10}$ &$1.67\pm0.08$&$13.30\pm2.23$&$8.48\pm0.97$&$1.30\pm0.06$&$2.65\pm0.27$ \\
%    $\rm SFR_{100}$ &$1.77\pm0.18$&$37.71\pm3.28$&$8.77\pm1.02$&$1.35\pm0.07$&$2.74\pm0.32$ \\
    $\rm M_{\star}$& $\rm M_\odot$&$(4.78\pm0.24)\times10^{11}$&$(3.83\pm0.79)\times10^{11}$&$(5.71\pm0.85)\times10^{11}$&$(2.14\pm0.27)\times10^{11}$&$(3.75\pm0.33)\times10^{11}$\\
    $\rm M_{\star,young}$& $\rm M_\odot$&$(1.62\pm0.08)\times10^{7}$&$(1.27\pm0.21)\times10^{8}$&$(8.21\pm0.95)\times10^{7}$&$(1.26\pm0.06)\times10^{7}$&$(2.55\pm0.26)\times10^{7}$\\
    $\rm M_{dust}$& $\rm M_\odot$&$(1.29\pm0.06)\times10^{8}$&$(4.87\pm0.27)\times10^{8}$&$(4.04\pm0.25)\times10^{8}$&$(1.33\pm0.07)\times10^{7}$&$(8.85\pm0.44)\times10^{7}$\\
    %$\rm M_{gas}$&$(2.01\pm0.10)\times10^{11}$&$(1.67\pm0.40)\times10^{11}$&$(2.39\pm0.43)\times10^{11}$&$(8.97\pm1.32)\times10^{10}$&$(1.67\pm0.16)\times10^{11}$\\
    $\rm L_{\star}$& $\rm L_\odot$&$(2.54\pm0.13)\times10^{11}$&$(2.96\pm0.15)\times10^{11}$&$(4.04\pm0.20)\times10^{11}$&$(1.22\pm0.06)\times10^{11}$&$(1.67\pm0.08)\times10^{11}$\\
    $\rm L_{dust}$& $\rm L_\odot$&$(4.72\pm0.24)\times10^{10}$&$(1.15\pm0.06)\times10^{11}$&$(1.05\pm0.05)\times10^{11}$&$(4.83\pm0.24)\times10^{9}$&$(2.16\pm0.11)\times10^{10}$\\
    $\rm A_{FUV}$& mag &1.017$\pm$0.003&$1.734\pm0.125$&1.632$\pm$0.034&0.857$\pm$0.003&0.865$\pm$0.011\\
    $\rm \tau_{main}$& Gyr &0.843&2.750&2.106&0.843&2.000\\
    $\rm \chi^{2}_{\nu}$ & $-$& 1.20 & 2.40 & 0.81 & 4.00 & 2.30 \\
    \hline\\[-1.5ex]
    $\rm \zeta$& $-$&$(2.95\pm0.21)\times10^{4}$&$(3.02\pm0.80)\times10^{3}$&$(6.95\pm1.31)\times10^{3}$&$(1.70\pm0.23)\times10^{4}$&$(1.47\pm0.20)\times10^{4}$\\
    $\rm M_{\star,peak}$& $\rm M_\odot$ &$1.73\times10^{11}(36.19\%)$&$1.62\times10^{11}(42.30\%)$&$2.30\times10^{11}(40.28\%)$&$0.66\times10^{11}(30.84\%)$&$1.25\times10^{11}(33.33\%)$\\
    $\rm M_{baryon}$& $\rm M_\odot$&$(5.10\pm0.24)\times10^{11}$&$(5.41\pm0.85)\times10^{11}$&$(5.99\pm0.85)\times10^{11}$&$(2.63\pm0.27)\times10^{11}$&$(4.11\pm0.34)\times10^{11}$\\
    $\rm M_{expected}$& $\rm M_\odot$&$2.81\times10^{12}$&$3.61\times10^{12}$&$1.82\times10^{12}$&$3.19\times10^{12}$&$2.66\times10^{12}$\\
    $\rm f_{b,r_{200}}$& $-$&$0.030\pm0.001$&$0.025\pm0.004$&$0.055\pm0.008$&$0.014\pm0.001$&$0.026\pm0.002$\\
    $\rm f_{\star}$& $-$&$0.170\pm0.009$&$0.106\pm0.022$&$0.314\pm0.047$&$0.067\pm0.008$&$0.141\pm0.012$\\
    \hline
    \end{tabular}
    \caption{Here, in the upper panel of the table, we mention the best-fit output parameters for SED fitting in CIGALE. The mentioned parameters are as follows: $\rm SFR_{inst}$ = instantaneous star formation rate, $\rm M_{\star}$ = stellar mass, $\rm M_{\star,young}$ = mass of young stellar population, $\rm M_{dust}$ = dust mass, $\rm L_{\star}$ = stellar luminosity, $\rm L_{dust}$ = dust luminosity, $\rm A_{FUV}$ = extinction in F154W FUV-band of UVIT, $\rm \tau_{main}$ = the peak age of star formation of the main stellar population of the objects, $\rm \chi^{2}_{\nu}$ = reduced chi-sqaure of the best fit. In the lower panel, we mention some more useful parameters calculated using SED-derived data. We show, $\rm \zeta = M_{\star,old}/M_{\star,young}$ = the ratio of the stellar mass of old to young population, $\rm M_{\star,peak}$ = stellar mass of the objects accumulated by time $\rm \tau_{main}$ with the percentage of the present day stellar mass $\rm M_{\star}$ in bracket, $\rm M_{baryon} = M_{\star} + M_{gas} + M_{dust}$ = total baryonic mass (see Table~\ref{tab:all_info} for $\rm M_{gas}$), $\rm M_{expected}$ = baryonic mass of the galaxies expected from the cosmic baryon fraction of 0.167 and the halo mass $\rm M_{halo}$ (see Table~\ref{tab:all_info}), $\rm f_{b,r_{200}} = M_{baryon}/M_{halo}$ = baryon fraction of the galaxies taking total baryonic mass and the halo mass upto the virial radius and, $\rm f_{\star} = M_{\star}/f_{b}M_{halo}$ = star formation efficiency.}
    \label{tab:sed_table}
\end{table*}
%%%%%%%%%%%%%%%%%%%%%%%%%%%%%%%%%

The star formation rates for the objects are found out following the relation of \cite{1998ARA&A..36..189K} between the SFR and far-ultraviolet luminosity $\rm L_{FUV}$ assuming a Salpeter Initial Mass Function (IMF) with mass limits 0.1$-$100 $\rm M_{\odot}$ which is given below,
%%%%%%%%%%%%%%%%%%%%%%%%%%%%%%%%%
\begin{equation}
\rm     SFR_{FUV} (M_\odot yr^{-1}) = 1.4\times10^{-28} L_{FUV} (erg\,s^{-1}Hz^{-1})
\end{equation}
%%%%%%%%%%%%%%%%%%%%%%%%%%%%%%%%%
Note that, we estimate all the SFRs in this paper considering a Salpeter initial mass function (IMF). The magnitude $\rm m_{FUV}$, SFR and $\rm \Sigma_{SFR}^{FUV}$ mentioned in Table~\ref{tab:astrodendro_stats} are corrected for both foreground and internal attenuation of the galaxies. To estimate the corrected magnitude and/or flux of the clumps, the effect of foreground attenuation of radiation is taken into account following the procedure described in Section~\ref{sec:photometry} and to eliminate the effect of the internal attenuation within the host galaxies in FUV wavelength, we consider the total extinction $\rm A_{FUV}$ (mag) found out from the best-fit spectral energy distribution of the objects using the grid shown in Table~\ref{tab:cigale_grid} and are mentioned in Table~\ref{tab:sed_table} (see Section~\ref{sec:sed_description} for details).

%\begingroup
%\begin{figure}
%    \centering
%    \includegraphics[width=0.5\textwidth]{ngc1961_fuv_structure_cut_4.eps}
%    \caption{Here, we show a cut-out region ($\rm \sim 57\times57$ arcsec$^{2}$) of star forming clumps (Figure~\ref{fig:clump_structures}) of the object NGC~1961. The green contours indicate the branches and the red contours indicate the child structures or the leaves. It can be seen that many of the leaves are isolated structures showing isolated regions of young star formation. Note that, the image pixel values are in the units of $\rm 1.87\times10^{-18}erg~s^{-1}~cm^{-1}~Hz^{-1}$ and here the values are scaled between 0 and 2.}
%    \label{fig:ngc1961_cutout_clumps}
%\end{figure}
%
%\begin{figure}
%    \centering
%    \includegraphics[width=0.5\textwidth]{dendrogram_structure_flux_ngc5635_fuv.eps}
%    \caption{Here, we show how the parent and child structures are distributed in the hierarchical star formation algorithm of ASTRODENDRO \citep{2019ascl.soft07016R}. We show a distribution of 40 structures out of 238 child structures identified in NGC~5635 (see Table~\ref{tab:astrodendro_stats}). The upper structures with the highest fluxes are called the child structures or the leaves (shown in the left graph) and all leaves are merging into few parent structures (can be seen from the right graph) also known as trunks that have no further parent structures. The flux mentioned here is in the units of $\rm 4.40\times10^{-19} erg~s^{-1}~cm^{-2}~Hz^{-1}$.}
%    \label{fig:dendrogram_ngc5635_fuv}
%\end{figure}
%\endgroup

\subsection{The Spectral Energy Distribution }\label{sec:sed_description}

To fit the spectral energy distribution (SED) of the objects, we have made use of data from our UVIT observations as well as other archival data for our objects as mentioned in Section~\ref{sec:observations}. The steps followed to estimate the fluxes in FUV-FIR bands of the objects are given in Section~\ref{sec:photometry}. We used CIGALE (v2022) (Code Investigating GALaxy Emission) \citep{2019A&A...622A.103B,2022ApJ...927..192Y} to perform  SED fitting, where we utilized multi-band photometry data of the objects given as inputs. The FUV$-$FIR fluxes are corrected for attenuation due to foreground dust before using them as inputs to CIGALE. The modules used to perform the fitting are described below,

\begin{enumerate}
    \item Star formation history: The module \textit{sfhdelayed} is used to model the star formation history of the objects. Here, the star formation model is defined as $\rm SFR(t) \propto \frac{t}{\tau^{2}}\exp{(-t/\tau)}$ for t as the variable of time and where $\rm \tau$ defines the peak of the star-forming history of a galaxy.
    \item Stellar populations: The module \textit{bc03} is used to model the stellar population of the galaxies \citep{2003MNRAS.344.1000B}.
    \item Nebular emission: The module \textit{nebular} is used to fit the nebular emission \citep{2011MNRAS.415.2920I} in galaxies beyond mid-infrared as a result of heating and ionisation of gas surrounding massive stars, which denotes young star-forming regions within the galaxy.
    \item Dust attenuation: The module \textit{dustatt$\_$calzleit} is used to model the attenuation of light in the galaxy. The dust absorbs radiation in shorter wavelengths (ultraviolet and near-infrared) and re-radiates them in mid and far-infrared. Here we use models as per \citet{2000ApJ...533..682C} to account for the attenuation.
    \item Dust emission: We use \textit{dl2014} to fit the re-processed radiation in mid and far-infrared and beyond \citep{2014ApJ...780..172D}.
    \item AGN: To account for the effect of the active galactic nuclei (AGN) in the overall SED, we make use of \textit{fritz2006} as per \citet{2006MNRAS.366..767F}.
\end{enumerate}

We have fitted the spectral energy distribution of each object in our sample individually. The CIGALE grid used to find out the best-fit SEDs for the objects is described in Table~\ref{tab:cigale_grid}, where we mention all the free parameters related to the models described above have been used to optimize the fittings. The final best-fit model SEDs for the objects and the observed multi-band photometric data points used as inputs are shown in graphs in Fig.~\ref{fig:sed_fitting}. This figure shows CIGALE's best-fit models of the objects' spectral energy distributions. The data points are shown with empty black circles, and the models are shown with black lines. 
%We have employed the following modules to estimate the best-fit results from the FUV-FIR SED: \textit{sfhdelayed} for star formation history SFH, \textit{bc03} for a single stellar population SSP, \textit{nebular} for nebular emission lines, \textit{dustatt$\_$calzleit} for dust attenuation, \textit{dl2014} for dust emission, \textit{fritz2006} for AGN contribution. 
In the lower parts of each graphs, we show the filled transmission curves for the bands used for wavelength references. Filters from left to right: UVIT F154W, UVIT N242W, SDSS u, SDSS g, SDSS r, SDSS i, SDSS z, WISE1, WISE2, WISE, WISE4, IRAS60, IRAS100. Note that we have also made use of DECaLS (g, r, z) and Pan-STARRS1 (g, r, i, z, y) images for NGC~1030 and NGC~1961 respectively (Section~\ref{sec:observations}).

As mentioned earlier, the free parameters for the fitting are shown in Table.~\ref{tab:cigale_grid} and it is important to note that, we have fitted each of the galaxies individually instead of a batch fitting in order to observe the variations of the parameters more carefully for each galaxies and change them accordingly to optimize the fitting though minimization of reduced chi-square ($\rm \chi_{\nu}^{2}$). At first, for each of our objects we start the fitting by the default script generated by CIGALE and then change the free parameters individually and take into account different combinations of them in order to obtain a better fit for each galaxies. This can be noted from the different set of choices for a single parameter (e.g. $\rm \tau_{main}$) in case of different galaxies. The free parameters corresponding to the modules described before in this section and also mentioned in the Table.~\ref{tab:cigale_grid} are described as the following: (i) star formation history is parameterized by e-folding time of the main stellar population model ($\rm \tau_{main}$), age of the main stellar population in the galaxy (Age), e-folding time of the late starburst population model ($\rm \tau_{burst}$), age of the late burst ($\rm Age_{burst}$) and mass fraction of the late burst population ($\rm f_{burst}$), (ii) Stellar population is parameterized by the initial mass function (IMF), metallicity, (iii) dust attenuation is parameterized by colour excess of the stellar continuum light for the young population ($\rm E(B-V)_{young}$), Reduction factor for the E(B-V) of the
old population compared to the young one ($\rm E(B-V)_{old-factor}$), (iv) dust emission is parameterized by mass fraction of Polycyclic Aromatic Hydrocarbon PAH ($\rm q_{PAH}$), power-law slope $\rm \alpha$ for $\rm \frac{dU}{dM}\propto U^{\alpha}$. Note that, here, we only mention the parameters that have been varied during the fitting (for more details on each module, see \citealt{2019A&A...622A.103B}).

The results of the SED fitting are tabulated in the upper panel of Table~\ref{tab:sed_table}. We mention the best-fit outputs of some of the parameters of our interest and their parent modules: (i) \textit{sfhdelayed:} instantaneous star formation rate $\rm SFR_{inst} (M_{\odot}yr^{-1})$, e-folding time of the main stellar population model $\rm \tau_{main} (Gyr)$, (ii) \textit{bc03:} total stellar mass $\rm M_{\star}$ ($\rm M_{\odot}$), stellar mass of the young population $\rm M_{\star,young}$ ($\rm M_{\odot}$), total stellar luminosity $\rm L_{\star}$ ($\rm L_{\odot}$), (iii) \textit{dustatt$\_$calzleit:} attenuation in UVIT FUV wavelength $\rm A_{FUV}$ (mag), (iv) \textit{dl2014:} dust mass $\rm M_{dust}$ ($\rm M_{\odot}$), dust luminosity $\rm L_{dust}$ ($\rm L_{\odot}$). Note that, in our best-fit SEDs of the objects, we found negligible/ almost no contribution from the AGN, indicating at the present day not much activity is taking place in the galactic nuclei of these massive spirals. In the lower panel of Table~\ref{tab:sed_table}, we show some more parameters estimated using SED-derived data and other parameters shown in Table~\ref{tab:all_info}. We estimate the old to young stellar population ratio $\rm \zeta$ from SED, where we find the mass of the older stellar population $\rm M_{\star,old}\sim M_{\star}$. The stellar mass accumulated up to the peak of the star-forming activity (denoted by $\rm \tau_{main}$) of the corresponding objects are denoted by $\rm M_{\star, peak} (M_{\odot})$. The total baryonic mass $\rm M_{baryon} (M_{\odot})$ of these galaxies are estimated using the stellar mass $\rm M_{\star}$, dust mass $\rm M_{dust}$ (both shown in upper panel, Table~\ref{tab:sed_table}) and gas mass $\rm M_{gas}$ (Table~\ref{tab:all_info}). The expected baryonic mass these spirals should have as per the mean cosmological baryon fraction is shown by $\rm M_{expected} = f_{b}M_{halo}$, where $\rm f_{b}=0.167$ and the halo mass $\rm M_{halo}$ is shown in Table~\ref{tab:all_info}. However, the baryon fraction that is seen in these galaxies within the virial radius $\rm r_{200}$ is denoted by $\rm f_{b,r_{200}}$ as the ratio of $\rm M_{baryon}$ to $\rm M_{halo}$. Finally, we mention the star formation efficiency of these objects, which is denoted by $\rm f_{\star} = M_{\star}/f_{b}M_{halo}$, which characterises the efficiency of halo baryons' conversion into stars. It can be seen that all of our spirals have stellar mass $\rm > 10^{11} M_{\odot}$ (and luminosity $\rm > 10^{11} L_{\odot}$) comparable to high-mass elliptical galaxies. The instantaneous star formation rate of these objects falls in between $\rm 1.29-12.04 M_{\odot}yr^{-1}$ considering a delayed star formation history model (Section~\ref{sec:sed_description}), where the peak star formation of these objects happened in between $\rm 0.843-2.750$ Gyr after the `Big Bang'.

The star formation rates of the objects as a function of the stellar mass is shown in (Fig.~\ref{fig:ssfr_mstar}, left panel). In this figure, the variation of specific star formation rate (sSFR in ${\rm yr^{-1}}$) versus the stellar mass ($\rm M_{\star}$ in \Msun) is  shown for galaxies of various types. The dotted line indicates the star-forming the main sequence from $z\sim0$ \citep{2007A&A...468...33E}. Highly Star-forming ultraviolet luminous galaxies are taken from \citet{2007ApJS..173..441H} (filled magenta circles). Super spirals, lenticulars and post-merger galaxies  are taken from \citet{Ogle_2019} (orange boxes with red edges).
%\citet{2017ApJS..233...20L} indicates the isolated massive spiral galaxies. 
A sample of galaxies with different quenching stages like the star-forming (filled blue circles), mixed (filled orange circles with black edges), nearly retired (filled red circles), fully retired (filled red circles with black edges) and quiescent-nuclear-ring (filled orange circles) taken from \citet[][]{2021A&A...648A..64K} have also been shown. The positions of the five target galaxies of the present study are shown with star symbols, with instantaneous SFRs and stellar masses given in Table~\ref{tab:sed_table}. Despite their high stellar masses, one can see that their sSFR is anomalously low, much below the main sequence, and close to the region of the most massive  super spirals of \citet{Ogle_2019}, but  above the  highly quenched, fully retired galaxies. The `Green Valley' region with $\rm 10^{-11.5}<sSFR (yr^{-1})<10^{-10.5}$ \citep{2016ApJS..227....2S} is shown with light green shadow.

Moreover, we also show the variation of the star formation efficiency $\rm f_{\star}$ of the galaxies as a function of its stellar mass $\rm M_{\star}$ in (Fig.~\ref{fig:ssfr_mstar}, right panel). Using a multi-epoch abundance matching model, the black curve is obtained from the relation between halo mass and the corresponding stellar mass provided by \citet{2013MNRAS.428.3121M}. The SPARC (Spitzer Photometry and Accurate Rotation Curves) sample of disc galaxies \citep{2016AJ....152..157L,2019A&A...626A..56P} is shown with black+grey points. In the case of $\rm M_{\star}/M_{halo} = f_{b}$, cosmic baryon fraction (CBF), the star formation efficiency $\rm f_{\star} = 1$ and is shown with the light grey line.

\section{Discussion}

\subsection{Star formation} \label{sec:sfr}

The ultraviolet radiation from the young star-forming regions of a galaxy gets absorbed by the dust in the interstellar medium (ISM) and ultimately re-emitted in the thermal infrared. \citet{1998ARA&A..36..189K} shows that the far-infrared luminosity integrated over $8-1000\micron$ is a sensitive tracer of young star formation in a galaxy. Now, to account for the absorbed FUV radiation by the dust in the ISM, we use the relation provided by \citet{2012ARA&A..50..531K} and estimate the corrected FUV luminosity using the following relation,

\begin{equation}
\rm    L^{corr}_{FUV} = L^{obs}_{FUV} + 0.46 L_{TIR}
\end{equation}

Here, $\rm L^{corr}_{FUV}$ and $\rm L^{obs}_{FUV}$ are the dust corrected and observed luminosity in far-ultraviolet, $\rm L_{TIR}$ = total infrared luminosity over $8-1000\micron$. We use the observed luminosity $\rm L^{obs}_{FUV}$ of the objects as per the foreground extinction corrected magnitudes mentioned in Table~\ref{tab:all_info}. The integrated luminosity is estimated as, $\rm L_{TIR}\sim 1.75 L_{FIR}$ \citep{2000ApJ...533..682C}, where the far-infrared luminosity $\rm L_{FIR}$ is found out using IRAS 60 and 100 $\micron$ fluxes following \citet{1988ApJS...68..151H}. The star formation rates of the objects using the dust-corrected FUV luminosities are shown in Table~\ref{tab:all_info} as $\rm SFR_{FUV}^{corr} (M_{\odot}yr^{-1})$. The spirals NGC~1961 and NGC~4501 have the highest FUV star formation of $\rm 13.66$ and $\rm 13.51 M_{\odot}yr^{-1}$ respectively followed by NGC~1030, NGC~266 and NGC~5635.

Massive galaxies often show a quenched state of star formation at low red-shift. The general idea has been that they acquire most of their mass earlier in time compared to the main-sequence galaxies. In Fig.~\ref{fig:ssfr_mstar} (left panel), we show the variation of the specific star formation rates of the galaxies as a function of their stellar masses for a wide range of samples, including galaxies at different stages of their evolution and properties where we mention ultraviolet luminous galaxies from \citet{2007ApJS..173..441H}, the super spiral, lenticular and post-merger galaxies from \citet{2019ApJ...884L..11O} and a mixed group of star-forming, nearly retired and fully retired galaxies from \citet{2021A&A...648A..64K}. It can be seen that our sample galaxies with stellar mass $\rm \gtsim 10^{11}M_{\odot}$ fall below the expectation of the main-sequence relation at $\it{z}\sim \rm 0$ \citep{2007A&A...468...33E}, specifically in the `Green Valley' region with specific star formation rates in $\rm 10^{-11.5}<sSFR<10^{-10.5} yr^{-1}$ \citep{2016ApJS..227....2S}, confirming the overall trend of lower star formation with increasing stellar mass which is evident from the graph (see Table~\ref{tab:sed_table} for $\rm M_{\star}$ and instantaneous SFR). Even if we consider the specific star formation rates of our objects using FUV dust-corrected star formation rates and their K-band stellar masses (see Table~\ref{tab:all_info} for $\rm sSFR_{FUV}^{corr} (yr^{-1})$), they fall in the `Green Valley' region indicating they have been in this state for at least the last $\rm \sim 100Myr$ considering FUV radiation is indicative of star formation in a galaxy in the past $\rm \sim 100Myr$ \citep{2012ARA&A..50..531K}. The intermediate spiral NGC~1961, which has the highest star formation in our sample, is the closest to the main-sequence relation, followed by NGC~4501. NGC~1030 shows the lowest instantaneous SFR, falling almost at the same place as J2345-0449, which is a spiral galaxy hosting a large-scale radio jet \citep{2014ApJ...788..174B}. However, our sample of late-type spirals shows a mix of moderate and/or nearly retired instantaneous star formation as opposed to a fully quenched state as shown by the massive isolated S0-a galaxy UGC~12591 \citep{2022MNRAS.517...99R}.

In Table~\ref{tab:sed_table} (upper panel), we mention the best-fit $\rm \tau_{main}$ for the objects, which indicates the peak of star formation activity in the galaxies according to a delayed SFR function as described in Section~\ref{sec:sed_description}. Based on our assumed star formation history, we find that all the objects barring NGC~1961 had experienced a peak in their star formation before the peak in the cosmic star formation history, the `cosmic noon' around $\rm 1<\it{z}\rm<3$ translating to a period of $\sim 2.2-5.9$ Gyr after the `Big Bang'. We also show that the objects already became massive by the time $\rm \tau_{main}$ and gained stellar mass $\rm M_{\star, peak}$ of the order of a $few \rm\times10^{11}M_{\odot}$ before and up to the peak of their star formation histories, corresponding to a growth of $\sim 31-42$ per cent in a time $\rm (1/16)$-$\rm (1/5)^{th}$ of the age of the Universe (Table~\ref{tab:sed_table}). According to our model, the objects that had experienced the peak relatively sooner after the `Big Bang' show a more suppressed state of star formation in recent times (e.g. NGC~1030, NGC~5635, NGC~266) compared to the objects that had their peak relatively in the later period (e.g. NGC~1961, NGC~4501) are still forming stars (Table~\ref{tab:all_info} and \ref{tab:sed_table}).

\begin{figure}
    \centering
    \includegraphics[width=0.5\textwidth]{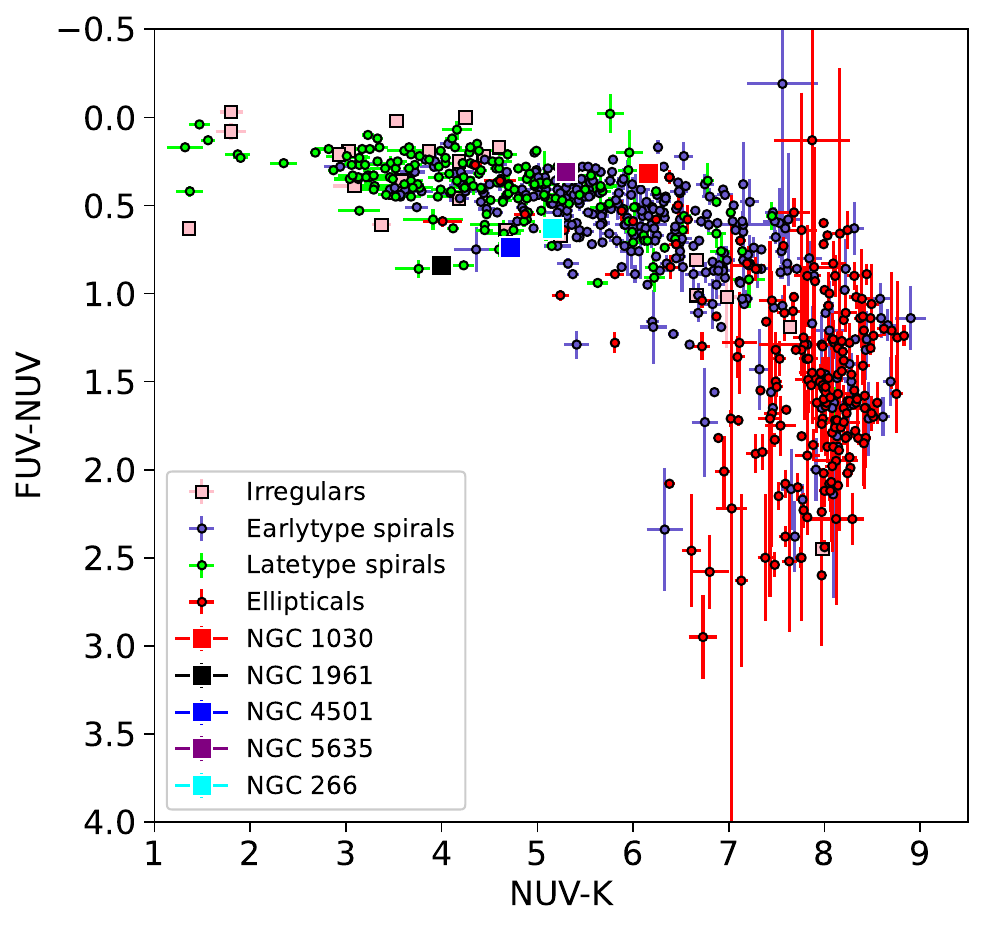}
    \caption{Here, we show the color-color diagram taking foreground extinction corrected (FUV-NUV) and (NUV-K) colors into account from the GALEX ultraviolet ATLAS of nearby galaxies \citep{2007ApJS..173..185G}. We plot our objects with galaxies of wide morphologies ranging from ellipticals (red) to late-type spirals (green).}
%    \caption{Here, we show the variation of the star formation rate density $\rm \Sigma_{SFR} (M_\odot yr^{-1} kpc^{-2})$ with the molecular and atomic mass densities, $\rm \Sigma_{H2}$ and $\rm \Sigma_{HI}$ in $M_\odot pc^{-2}$ respectively.}
    \label{fig:color_color}
\end{figure}

From the spectral energy distribution of these late-type spirals, we also found the contribution of the older stellar population and the younger stellar population in determining the total stellar mass of these objects. We mention the stellar mass of young population $\rm M_{\star, young}$ in Table~\ref{tab:sed_table} (upper panel), and it can be seen that these are much smaller than the total stellar mass of the objects, i.e. $\rm M_{\star, young}<<M_{\star}$. However, NGC~1961 and NGC~4501, having the highest young star formations among the five objects, have $\sim10$ times more young stellar populations than the other three spirals. We find that these massive galaxies are dominated by older stars, as can be seen from the ratio $\rm \zeta$ between old and young stellar mass in Table~\ref{tab:sed_table} (lower panel) with $\rm M_{\star, old}\sim M_{\star}$. This is also evident from the fact that for our objects K-band stellar mass $\rm M_{K}\sim M_{\star}$ (see Table~\ref{tab:all_info}, Section~\ref{sec:baryons}), considering K-band luminosity of a galaxy is primarily dominated by radiations from the galaxy's older stellar population. And objects having a higher population of older stars relative to young stars (hence higher value for $\rm \zeta$) show lower $\rm \tau_{main}$ (Table~\ref{tab:sed_table}) as their main stellar population had peaked sooner after the `Big Bang'. This is also visible from the graph in Fig.~\ref{fig:color_color}, where we show foreground extinction corrected color-color diagram (considering FUV-NUV and NUV-K colors) for elliptical, spiral and irregular galaxies in the nearby Universe from \citet{2007ApJS..173..185G}. In that graph, we find the color NUV-K for our objects increases as $\rm \tau_{main}$ decreases indicating a relatively more older stellar population in objects with smaller $\rm \tau_{main}$, evident from our discussion about $\rm \zeta$ and $\rm \tau_{main}$ (Table~\ref{tab:sed_table}). Now, as per the hierarchical structure formation theory of the standard $\rm \Lambda$CDM cosmology, where it is assumed that massive systems like these galaxies are formed through the merger of smaller objects over some time, so, they must have more young stellar population instead of older stellar population. The lack of understanding of this well-known problem related to this aspect of galaxy formation theory has also been reported by \citet{2018NatAs...2..695M}, \citet{2005ApJ...621..673T,2010MNRAS.404.1775T}. One possible solution to that problem could be that these massive galaxies are formed through merging with relatively more minor objects with their own rapid `in situ' star formations that occurred in the early universe, and the merged system undergoes negligible new star formation in the later period of evolution as suggested by \citet{2010ApJ...725.2312O}. However, careful investigation of the high redshift galaxies, more rigorous parameterization of the star formation histories, and detailed understanding of the baryonic processes in the dark matter halos are some of the key aspects to exactly establish and perhaps solve the problem.

However, the recent periods ($\rm \sim 100Myr$) of star formation seen predominantly in NGC~1961 and NGC~4501 are most probably due to some recent interaction(s) that caused asymmetries in the galactic plane, clearly visible in NGC~1961 optical image as we discussed in Section~\ref{sec:sample_selection}. Even though there are no signatures of recent head-on mergers in our objects, the turbulence in the galactic disc provided by strong interaction(s) can make it susceptible to new star formations \citep{2011EAS....51..107B}. In Section~\ref{sec:sample_selection}, we mention that all our objects show a varying order of asymmetry and/or interaction, which has helped them to revive star formation to some degree, unlike the quenched disc galaxy UGC~12591 \citep{2022MNRAS.517...99R} situated in an isolated environment devoid of any interaction. However, the exact mechanism of the process is contested \citep{2019A&A...631A..51P} and may vary if subjected to morphologically different galaxies and the environment they reside in. Our results are in parallel to the findings of \citet{2020ApJ...895..100X} who show that although the massive galaxies are mostly quiescent in nature, there exists a fraction ($\rm \sim 20$ per cent) of them that are forming stars and of them, $\rm \sim 85$ per cent have asymmetries induced in their structures by recent mergers.

\subsection{Star forming clumps}

The UVIT far-ultraviolet images have been spatially resolved into hierarchical star-forming clumps and are shown in Fig.~\ref{fig:clump_structures} (First row). We show the statistics of the identified clumps in the consequent rows of Fig.~\ref{fig:clump_structures}. We find the least number of structures in NGC~1030 (Table~\ref{tab:astrodendro_stats}) corresponding to very few regions of ongoing star formation with a radius ranging between 160-317 pc partly due to the edge-on view and the existence of dust lanes as can be seen from the optical image Fig.~\ref{fig:optical_images}. The identified clumps vary over a magnitude of $\sim$ 1-mag showing similar star-formation going on in them with mean star formation rate (SFR) of $\rm 0.0177\pm0.0078 M_{\odot}yr^{-1}$ and mean star formation rate density ($\rm \Sigma_{SFR}^{FUV}$) of $\rm 0.038\pm0.008 M_{\odot}yr^{-1}kpc^{-2}$. In NGC~1961 we find 393 clumps (Table~\ref{tab:astrodendro_stats}) with magnitude varying over $\sim$ 3-mag and having the highest SFR and $\rm \Sigma_{SFR}^{FUV}$ with mean value of $\rm 0.0110\pm0.0079 M_{\odot}yr^{-1}$ and $\rm 0.098\pm0.034 M_{\odot}yr^{-1}kpc^{-2}$ respectively. Even though the dust-corrected FUV SFR is similar for NGC~1961 and NGC~4501 (Table~\ref{tab:all_info}), their clumps statistics differ significantly. NGC~4501 has the highest number of star-forming clumps, with 1511 clumps having a radius between 34-168 pc and mean SFR and $\rm \Sigma_{SFR}^{FUV}$ of $\rm 0.0024\pm0.0020M_{\odot}yr^{-1}$ and $\rm 0.066\pm0.024 M_{\odot}yr^{-1}kpc^{-2}$ respectively. The clumps of the object NGC~266 show lower mean SFR and $\rm \Sigma_{SFR}^{FUV}$ of $\rm 0.0021\pm0.0014 M_{\odot}yr^{-1}$ and $\rm 0.014\pm0.003M_{\odot}yr^{-1}kpc^{-2}$ respectively followed by NGC~5635 with the parameters as $\rm 0.0017\pm0.0016M_{\odot}yr^{-1}$ and $\rm 0.011\pm0.004M_{\odot}yr^{-1}kpc^{-2}$ respectively. It can be seen that the mean local dust corrected magnitude $\rm m_{AB}$ and star formation rate density $\rm \Sigma_{SFR}^{FUV}$ of the clumps tend to vary proportionately with the global dust corrected SFR $\rm SFR_{FUV}^{corr}$ (Table~\ref{tab:all_info}) of the corresponding objects rather than the local mean dust corrected SFR of the clumps (Table~\ref{tab:astrodendro_stats}). We have spatially constrained the extent of star-forming regions in these galaxies, which is evident from the variation of clump SFR ($\rm M_{\odot}yr^{-1}$) with radius $\rm r_{d}$ ($\rm kpc$) in Fig.~\ref{fig:clump_structures} (Last row). NGC~1961 shows a maximum extent of star formation up to $\rm \sim 50 kpc$ with only a few structures after $\rm \gtrsim 30 kpc$, and on the other extent, NGC~1030 shows very few structures extending only up to $\rm \sim 12kpc$. The peaks of the clump SFR variation can be seen from the figure denoting spikes in the number of clumps in the spirals arms of the galaxies, which is more evident for objects like NGC~1961 and NGC~266 due to their face-on view.

It can be seen that the young star-forming clumps in these massive galaxies that are spatially resolved through UVIT FUV images show a local star formation rate density $\rm \Sigma_{SFR}^{FUV}$ varying over an order or 10, ranging from approximately $\rm 10^{-2}-10^{-1} M_{\odot} yr^{-1}kpc^{-2}$ (Table~\ref{tab:all_info}, Fig.~\ref{fig:clump_structures}) even though their global star formation rate density considering the corrected FUV star formation within aperture mentioned in Table~\ref{tab:all_info} varies in the range $\rm \sim 10^{-3.7}-10^{-2.7} M_{\odot} yr^{-1}kpc^{-2}$ signify the fact that the star formation in these massive spirals are highly localised with young star formation going on only locally throughout the galaxy similar to the giant SB0/a Low Surface Brightness (LSB) galaxy Malin1 \citep{2021JApA...42...59S}. This can further be investigated in detail with sensitive H$\alpha$ observations of these objects.

\subsection{Stellar and Interstellar components}\label{sec:baryons}

The less dust-affected near-infrared K-band luminosity can be taken as a proxy for the stellar mass of galaxies. Estimating the mass-to-light ratio $\rm \Upsilon^{\star}_{K}$ of the objects using \cite{2003ApJS..149..289B} based on their color (B-V) (we use u-g in the absence of B-V for NGC~1030 and NGC~5635), the galaxies are found to be extremely massive with stellar masses of the order of a $few  \times 10^{11} M_\odot$. The masses $\rm M_{K}$ and the corresponding K-band mass-to-light ratios $\Upsilon^{\star}_{K}$ (each $<1$) are shown in Table~\ref{tab:all_info}. It is also important to note that the K-band stellar masses show approximately the same results as the masses calculated from the spectral energy distribution of the objects (Table~\ref{tab:sed_table}, Fig.~\ref{fig:sed_fitting}).

%, which is considered to be more reliable as it rules out any biases arising from color dependence.

The interstellar medium (ISM) of a galaxy consists of atomic $\rm HI$, molecular $\rm H_{2}$ gas components, and dust. Hydrogen gas components in the ISM can be thought to be enveloping the young star-forming regions, with ionised hydrogen gas existing in the inner region surrounding the star-forming cloud and the atomic and molecular hydrogen gas occupying the middle and outer regions, respectively. To calculate the atomic HI gas mass, we consider the HI integrated line flux from the 21 cm line profile (\citealt{1988ApJS...67....1T,2005ApJS..160..149S,2011AJ....142..170H}) of the objects. The atomic gas mass $\rm M_{HI}$ can be calculated using the following relation,

\begin{equation}
\rm    M_{HI} = 2.356\times10^{5}\times D^{2}\times(1+\it{z})^{-2}\times \int S_{\nu}d\nu
\end{equation}

Here, D is the distance to the object in Mpc, $\it{z}$ is the redshift, and $\rm \int S_{\nu}d\nu$ is the integrated line flux in $\rm Jy\,km\,s^{-1}$.
The galaxy HI gas masses are found to be $\rm \approx 10^{10} M_\odot$ for our objects. The HI gas mass is tabulated in Table~\ref{tab:all_info}.

Now, to estimate the molecular gas content of our objects, we use the morphology-dependent relation between the molecular and atomic gas mass $\rm (M_{H2}/M_{HI})$ provided by \citet{1989ApJ...347L..55Y}. Depending on the morphological classifications of the objects as Sa to Sc types \citep{2014A&A...570A..13M} we estimate the molecular gas mass to be around $\rm \sim 10^{10} M_{\odot}$ (Table~\ref{tab:all_info}) for the objects. Note that, here we have assumed that the molecular gas content of an intermediate bar type (SABb) spiral and a barred Sab spiral does not differentiate much from a non-barred Sab spiral to calculate the molecular gas content for the object NGC~1961 and NGC~266. The total gas mass of the objects is then calculated as $\rm M_{gas} = 1.38\times(M_{HI}+M_{H2})$.

The dust mass for the objects is calculated from the spectral energy distribution (Fig.~\ref{fig:sed_fitting}) using the model provided by \citet{2014ApJ...780..172D}. The estimated mass $\rm M_{dust}$ for the objects fall between $\rm \approx 10^{7-8} M_{\odot}$ (Table~\ref{tab:sed_table}). The total baryonic mass ($\rm M_{baryon} = M_{\star} + M_{gas} + M_{dust}$) of the objects and their baryonic fraction $\rm f_{b,r_{200}}$ upto the virial radius are mentioned in Table~\ref{tab:sed_table} (Lower panel). It can be seen that the baryon fraction of these objects up to the virial radius varies from $\rm \sim 0.014-0.055$ indicating $\rm \sim 67-92$ per cent fewer baryons (`missing baryons') than expected (see Table~\ref{tab:sed_table} for $\rm M_{expected}\sim 10^{12}M_{\odot}$) according to the cosmic baryon fraction of 0.167. The most significant explanation of these `missing baryons' has been answered by the X-ray detection of hot gas around these massive galaxies that contribute significantly to the total baryonic mass. Now, if we consider the isothermal (constant temperature) profile of the halo gas, then the virial temperature $\rm T_{vir}$ of the halo gas as a function of the flat circular rotation velocity $\rm v_{c}$ can be expressed as,

\begin{equation}
 \rm   T_{vir} = 35.9\times\left(\frac{v_{c}}{km\,s^{-1}}\right)^{2} K
\end{equation}

Using the circular velocities mentioned in Table~\ref{tab:all_info}, we estimate the halo gas temperatures to be of the order of a $few \times10^{6}$K translating to $\rm 0.3-0.5 keV$ (Table~\ref{tab:all_info}) for our objects.

According to \citet{2021MNRAS.502.2934K}, the hot gas ($\rm > 5\times10^{5} K$) starts to dominate the total baryon fraction of a galaxy for virial mass beyond $\rm 10^{12}M_{\odot}$, who investigated the X-ray gas surrounding galaxies with halo mass in the range $\rm 10^{11}-10^{14}M_{\odot}$ in the cosmological EAGLE simulations. If we consider the virial mass of our objects ($\rm \sim 10^{13}M_{\odot}$), then the ratio of the hot gas to stellar mass component becomes $\rm \sim 5$ at red-shift $\rm \it{z}=$ 0 (see Figure~1; \citealt{2021MNRAS.502.2934K}), which gives us hot gas mass around $\rm \sim 10^{12}M_{\odot}$ for the stellar mass of our objects (Table~\ref{tab:sed_table}) solving the `missing baryon' problem keeping in mind uncertainties in individual parameters. But this vast amount of hot gas has not been seen in massive spirals due to the low density of the hot gas limiting X-ray observations to $\rm \sim 1/5$th of the virial radius \citep{2021MNRAS.500.2503M}. In our sample of massive spirals, out of the five objects, two have reported detections of hot ($\rm \sim10^{6}K$) X-ray halos extending much beyond their optical radius and/or the stellar disc. \citet{2013ApJ...772...97B} detected hot gas emission up to $\rm \sim 60$ kpc for NGC~1961 based on XMM-Newton X-ray observations with the hot halo gas mass of $\rm \sim (1.2\pm0.2)\times10^{10}M_{\odot}$ and in \citet{2013ApJ...772...98B} they report hot gas emission upto $\rm \sim 70$ kpc containing a mass of $\rm \sim (9.1\pm0.9)\times10^{9}M_{\odot}$ using ROSAT and Chandra X-ray observations of NGC~266. These detections give us a hot halo gas mass of 2 orders of magnitude less than what is expected ($\rm \sim 10^{12} M_{\odot}$). The high temperature ($\rm \gtrapprox 10^{6}K$) (Table~\ref{tab:all_info}) and the low-density halo gas significantly affect the star formation in the host galaxy by not being able to settle down in the galactic disc at temperature $\rm < 10^{2}K$ necessary for forming molecular clouds capable of forming stars.

\subsection{Black hole and its effect on star formation} \label{sec:bh_sfr}

The massive galaxies are often hosts to supermassive black holes at their centres \citep{1998AJ....115.2285M,2000ApJ...539L..13G}. It is extremely difficult to constrain the black hole masses of galaxies observationally. So, in order to estimate the black hole mass of the target objects, we use the tight correlation between the mass of the black hole $\rm M_{bh}$ and the central velocity dispersion $\rm \sigma$ of the spirals \citep{2009ApJ...698..198G}. The relation is shown below,
\begin{equation}
\rm    log(\frac{M_{bh}}{M_\odot}) = (8.12\pm0.08) + (4.24\pm0.41)log(\frac{\sigma}{200 km\,s^{-1}})
\end{equation}

The central velocity dispersion $\rm \sigma$ of the objects is mentioned in Table~\ref{tab:all_info}. In case the information about the central velocity dispersion is not available for the objects, such as the case for NGC~1030 and NGC~5635, we estimate the black hole mass $\rm M_{bh}$ from the relation provided by \citet{2015ApJ...813...82R} correlating the black hole mass with the stellar mass (see Table~\ref{tab:sed_table} for stellar mass). From these relations, we show that the black hole mass of our objects is in the range of a $few\rm\times10^{7-8} M_\odot$ and is tabulated in Table~\ref{tab:all_info}.

The star formation in massive galaxies ($\geq 10^{11}M_\odot$) is still and primarily an unexplored phenomenon; galaxies of such types can be broadly classified into two groups where a group of galaxies showing extremely quenched SFRs and the other group showing a moderate star formation (Fig.~\ref{fig:ssfr_mstar}, Left panel) for a similar mass range. The massive galaxies are supposed to have acquired much of their stellar mass content before normal main-sequence spirals, which are still actively forming stars, which also determined their growth of central black hole at a faster rate, which is evident from the study carried out by \citet{2015ApJ...813...82R} on a sample of 341 galaxies including galaxies hosting active galactic nuclei (AGN) and inferred that, the growth of black hole mass is accompanied by the growth of stellar mass of the host galaxy, i.e. $\rm M_{\star} \propto M_{bh}$. As we have discussed in Section~\ref{sec:sfr} that, depending on our assumed models, these massive galaxies have experienced their peak in the star-forming activity approximately in the period of red-shift $\it{z}\rm\sim2.4-6.5$ (Table~\ref{tab:sed_table}), reaching up to the `cosmic high noon' at $\rm \it{z}\rm\sim1-3$, which overlaps with the `quasar epoch' around red-shift $\rm \it{z}\rm\sim2$. This suggests the coexistence of the era of intense star formation with the `quasar epoch' in the universe \citep{1999ASPC..156..163S}. However, from the star formation history of our objects, it is evident that there existed a period of rapid star formation assisted by the black holes' growth as well as rapid stellar mass growth in the pre-quasar era leading up to the `quasar epoch' \citep{2005Natur.434..738A}. It indicates the possibility that the rise of star formation in these galaxies and the steep growth of the central black hole mass stops at around the peak of the star formation history of the galaxies, as by then, the black hole must have experienced a growth upto $\rm \sim10^{7}-10^{8} M_{\odot}$ to be able to regulate star formation of the host galaxies efficiently.

Now, we discuss the possible growth scenario of the central black holes in these massive galaxies in their early period of evolution. If the black hole mass at time t is $\rm M_{bh}(t)$, then the dimensionless Eddington luminosity ratio $\rm \lambda (=L/L_{Edd})$ can be defined as \citep{Shapiro_2005,Hopkins_2006},

\begin{equation}
    \rm  \lambda = \frac{\epsilon_{r} t_{sal}}{t (1-\epsilon_{r})}~log_{e}\frac{M_{bh}(t)}{M_{bh}(0)}
\end{equation}

Here, $\rm M_{bh}(0)$ is the initial black hole mass at time $\rm t=0$, $\rm \epsilon_{r} = 0.06-0.42$, for spin $\rm 0-1$ respectively, denotes how efficiently can the accreted mass be converted into radiative energy, $\rm t_{sal} = 0.45$ Gyr, the Salpeter timescale. For a scenario of the rapid growth of stellar mass (hence high SFR) and $\rm M_{bh}$, we expect the radiative efficiency $\rm \epsilon_{r}$ to be smaller as higher $\rm \epsilon_{r}$ will lead to lower growth rate for the black hole mass following above equation, $\rm dM_{bh}/dt \propto (1/\epsilon_{r} - 1)$. In our calculations we consider, $\rm \epsilon_{r}\sim0.1$ and $\rm M_{bh}(0)$ to be $\rm 10^{2}M_\odot$, a Population-III seed black hole. If we consider the central black holes of our spirals to have grown up to $\rm \sim 10^{8}M_\odot$ around the peak of their consequent star formations, then we estimate the ratio $\rm \lambda$ to be in the range $\sim \rm 0.25-0.86$ for $\rm \tau_{main} \sim 2.8-0.8$ Gyr (Table~\ref{tab:sed_table}), i.e. with $\rm \lambda \sim (1/\tau_{main})$ the accretion approaches the near-Eddington limit ($\rm \lambda\sim1$) for galaxies experiencing star forming peak sooner ($<10^{9}$ yr) after the `Big Bang'. So, It can be seen that the black hole growth and the star formation in a galaxy are closely related to each other, and our results follow the findings of \citet{2018Natur.553..307M}, who suggest that the black hole growth in the early universe is proportional to the gas cooling rate and hence the star-forming activity and galaxies hosting more massive black holes experience the suppression of star formation earlier than the others in their later period of evolution. However, from the SED fitting of these objects, we find almost no activity of the active galactic nuclei of these objects in the present time, which is also evident from the AGN class of the objects as mentioned in Table~\ref{tab:all_info}. As we do not see any large-scale radio jets from the centre of these objects, we infer that these AGNs are presently in radio-quiet mode, with possibly weak and small-scale outflows, with no significant effect on the star formation in the galactic disc.

\subsection{Star forming efficiency} \label{sec:sfe}

How efficiently the hot halo baryonic mass is converted into the stellar mass of a galaxy can be characterised by star formation efficiency. The star formation efficiency SFE of a galaxy is defined as the following \citep{2019A&A...626A..56P}, 

\begin{equation}
\rm    f_{\star} = \frac{M_{\star}}{f_{b} M_{halo}}
\end{equation}

Here, $\rm M_{\star}$ and $\rm M_{halo}$ are the stellar and halo mass of a galaxy, whereas $\rm f_{b}$ is the cosmic baryon fraction taken as 0.167 in our calculations \citep{2011ApJS..192...18K}. To calculate SFE, we use the stellar masses from the best-fit SEDs of the objects (Table~\ref{tab:sed_table}), and the halo masses of our objects are estimated using the following relations,

\begin{equation}
 \rm   M_{halo} = \frac{4}{3}\pi r_{vir}^{3}\Delta_{c}\rho_{c};~where,~r_{vir} = \frac{GM_{halo}}{v^{2}_{rot}}~\&~\rho_{c} = \frac{3H^{2}(\it{z})}{8\pi G}
 \end{equation}
 
Here, we assume that the density of the halo is $\rm \Delta_{c}\rho_{c}$ with $\Delta_{c}\sim200$ and $\rm \rho_{c}$ as the critical density of the universe. Under the assumption of flat circular rotational velocity $\rm v_{rot}$ up to the virial radius $\rm r_{vir}$, the halo mass can be written as, $\rm M_{halo} = \frac{v_{rot}^{3}}{10GH(\it{z})}$; G = Gravitational constant and $\rm H(\it{z})$ = Hubble parameter. Expressing the redshift dependent Hubble parameter as, $\rm H(\it{z}) = \rm H_{0}[\Omega_{M}(1+\it{z})\rm^{3} + \Omega_{vac}]^{1/2}$, the halo masses are estimated as $few\times10^{13}~\rm M_\odot$ (see Section~\ref{sec:introduction} for $\rm H_{0}, \Omega_{M}$ and $\rm \Omega_{vac}$). The estimated halo masses and SFEs are shown in Table~\ref{tab:all_info} and Table~\ref{tab:sed_table}, respectively. The star-forming efficiency of our sample of the massive spirals is shown in (Fig.~\ref{fig:ssfr_mstar}, right panel) along with other massive spirals like UGC~12591 \citep{2022MNRAS.517...99R} and J2345-0449 \citep{2014ApJ...788..174B,2021A&A...654A...8N}. We also show the SPARC (Spitzer Photometry and Accurate Rotation Curves) sample of disc galaxies from \citet{2016AJ....152..157L} having stellar mass in the range of $\rm \sim 10^{7}$ to $\rm > 10^{11} M_{\odot}$ in the graph. The star formation efficiency of the galaxies is supposed to peak around the stellar mass of $\rm M_{\star}\sim10^{10.2-10.3}~\rm M_\odot$ with $\rm f_{\star}\sim20$ per cent \citep{2013MNRAS.428.3121M}. This means that massive galaxies in their evolution towards the stellar mass of $\rm \sim 10^{11} ~\rm M_{\odot}$ can only experience a maximum $\rm \sim20$ per cent SFE with most of their baryons not converted into stars. However, the massive spirals from \citet{2016AJ....152..157L,2019ApJ...884L..11O}, as shown in the graph, depict a contradictory trend at the high mass end where the SFE increases with increasing stellar mass upto $\rm > 10^{11}~\rm M_\odot$ and corresponding $\rm f_{\star}$ reaching $\approx 0.3-1$ indicating the conversions of the most ($\rm 30-100$ per cent) of their baryons into stars. In contrast to this, an extremely massive, quenched and isolated spiral like UGC~12591 shows an SFE of $\sim 3-5$ per cent and falls in the place expected from \citet{2013MNRAS.428.3121M} model for red-shift $\it{z}\rm\sim0$. Our sample set of massive spirals with even greater stellar masses than the other samples mentioned before shows a star-forming efficiency of $\sim 7-30$ per cent with NGC~4501 and NGC~5635 having the highest and the lowest $\rm f_{\star}$ respectively (Table~\ref{tab:sed_table}) indicating a moderate to low ongoing ($\it{z}\rm\sim0$) conversions of baryons into stars. As we have discussed before in Section~\ref{sec:sfr}, our objects fall in the `Green Valley' region considering some signatures of recent star formation as evident from Fig.~\ref{fig:ssfr_mstar} (left panel) arguably induced by interactions. However, they fall below the star formation efficiency line for $\rm f_{\star} =1$, the line showing all baryons have been converted into stars. Moreover, they have higher SFEs than expected from the stellar-to-halo mass relation provided by \citet{2013MNRAS.428.3121M} by abundance matching model (Fig.~\ref{fig:ssfr_mstar}, right panel). This indicates the possibility that the stellar masses of these objects may have grown recently due to turbulence-induced star formation in the disc and possible gas inflow to some extent; however, large-scale baryon cooling and/or halo gas ($\rm \sim 10^{6}K$) condensation in the galactic disc is not very predominant provided their baryon fractions $\rm f_{b,r_{200}}$ (Table.~\ref{tab:sed_table}) are much less than the cosmic baryon fraction.

\section{Conclusions}

In this paper, we study the aspects related to star formation in massive spirals ($\rm > L_{\star}$) with a focus on five late-type galaxies NGC~1030, NGC~1961, NGC~4501, NGC~5635 and NGC~266 observed in UltraViolet Imaging Telescope onboard ASTROSAT. In addition to analysing the UVIT data, we used the archival data of GALEX, SDSS, PanSTARRS, DECaLS, 2MASS, WISE and IRAS in our analysis. The significant results of our analysis are listed below:

\begin{enumerate}
    \item The late-type spiral galaxies show a mix of moderate to nearly quenched young ($\rm \sim 100$ Myr) FUV star formation. Taking dust correction into account NGC~1961 show the highest star formation of $\rm SFR_{FUV}^{corr} = 13.66\pm2.04$ followed by NGC~4501 with $\rm SFR_{FUV}^{corr} = 13.51\pm2.24 M_{\odot} yr^{-1}$ and the other three galaxies NGC~1030, NGC~5635 and NGC~266 show $\rm SFR_{FUV}^{corr} = $ $\rm5.15\pm0.87, 1.43\pm0.13$ and $\rm2.54\pm0.25$ $~\rm M_{\odot} yr^{-1}$ respectively.\\

    \item We identify the young star-forming regions in the galaxies within a defined aperture, considering them to be hierarchical in nature and segregating them into clumps. We find that the star formation in these galaxies is highly localised with star formation rate density $\rm \Sigma_{SFR}^{FUV}$ in the range of $\rm\sim 10^{-2}-10^{-1}M_{\odot}yr^{-1}kpc^{-2}$ as opposed to a global star formation rate density of $\rm \sim 10^{-3.7}-10^{-2.7}M_{\odot}yr^{-1}kpc^{-2}$.\\
    
    \item From the  FUV-FIR multi-wavelength SED fitting of each of the galaxies, we find the stellar mass $\rm M_{\star}$ of the objects to be of the order of a $ few\times10^{11} \rm M_{\odot}$ with instantaneous star formation falling in the `green valley' region with $\rm 10^{-11.5}<sSFR (\rm M_{\odot} yr^{-1})<10^{-10.5}$ and with $\rm SFR_{inst}$ in the range $\rm \sim 1.7-12.0~\rm M_{\odot}yr^{-1}$. We find no significant AGN contributions affecting the star formation of these objects in the present era. However, all our objects show a varying degree of asymmetry/interactions, which most probably explains the recent star formation activity in these objects.\\
    
    \item Based on a delayed star formation history model, the peak of the star formation history of the objects is found to be in the range $\rm \sim 0.8-2.8$ Gyr after the `Big Bang' expanding throughout pre-cosmic `high noon' and pre-quasar era ($\rm 1<\it{z}\rm<3$). The galaxies that had experienced peaks sooner after the `Big Bang' show relatively low star formation at low redshift ($\it{z}\rm \sim 0$) than the others. For example, NGC~1030 and NGC~5635 experienced their peak $\rm \sim 0.843$ Gyr after the `Big Bang' have low instantaneous star formation rates of $\rm 1.66\pm0.08$ and $\rm 1.29\pm0.06 ~\rm M_{\odot}yr^{-1}$, whereas, NGC~1961 shows a relatively higher instantaneous star formation of $\rm 12.04\pm2.48 ~\rm M_{\odot}yr^{-1}$ having experienced the peak $\rm \sim 2.75$ Gyr after the `Big Bang'.\\

    \item We show that the objects had gained much of their stellar mass ($ > 10^{11} \rm M_{\odot}$) very rapidly in the early period of their evolution, acquiring almost $\rm \sim 31-42$ per cent of the present-day stellar mass in a period of $\rm \sim (1/16)-(1/5)^{th}$ of the age of the Universe. We argue that the early period of growth of the central black holes in these objects occurred rapidly with star formation and stellar mass growth in these objects. We show that the black holes must have accreted at near Eddington limit for objects that had their peak in the star formation at around $\rm \ltsim 10^{9}yrs$ after the `Big Bang' to be able to grow up to $\rm 10^{8}M_{\odot}$ and affect star formation activity of the host.\\

    \item We find that these spirals have baryonic mass (a $few \times\rm 10^{11}M_{\odot}$) of $\rm \sim 67-92$ per cent less than what is expected (a $few \times\rm 10^{12}M_{\odot}$) from the cosmic baryon fraction of 0.167. X-ray hot halo mass detected in two objects, NGC~1961 and NGC~266, tries to compensate for the `missing baryons'; however, they still are far less than what is expected. These objects have star-forming efficiency, the baryon-to-star conversion efficiency in the range $~\rm \sim 7-31$ per cent with halo mass of $\rm > 10^{13} ~\rm M_{\odot}$ and hot halo gas temperature of $\rm \sim 10^{6}$ K. With baryon fraction $\rm f_{b},r_{200}$ far less than 0.167, we conclude that there is no large scale baryon cooling happening in these extremely massive objects.\\

\end{enumerate}

\section*{Acknowledgements}

The authors are grateful to the anonymous referee for the encouraging and constructive comments on the manuscript, which greatly helped us in improving its quality. This publication uses the data from the UVIT, which is part of the ASTROSAT mission of the ISRO, archived at the Indian Space Science Data Centre (ISSDC). We thank Dr. Kanak Saha for his help during the observation proposal writing of target galaxies to the ASTROSAT. SR gratefully acknowledges the support from the Department of Physics and Electronics, CHRIST (Deemed to be University), Bangalore and funding by the Indian Space Research Organisation (ISRO) under ‘AstroSat Data Utilization’ project. SD acknowledges support from the Indian Space Research Organisation (ISRO) funding under project PAO/REF/CP167. JB acknowledges the support from the Department of Physics and Electronics, CHRIST (Deemed to be University), Bangalore. MBP gratefully acknowledges the support from the following funding schemes: The Science and Engineering Research Board (SERB), New Delhi, under the SERB ‘SERB' Research Scientists Scheme Scheme and Indian Space Research Organisation (ISRO) under the ‘AstroSat Data Utilization’ project. MBP also acknowledges the support from IUCAA Associateship programme.  We acknowledge the usage of the HyperLeda database (http://leda.univ-lyon1.fr). This research has made use of NASA’s Astrophysics Data System and of the NASA/IPAC Extragalactic Database (NED), which is operated by the Jet Propulsion Laboratory, California Institute of Technology, under contract with the National Aeronautics and Space Administration. This research has made use of the VizieR catalogue access tool, CDS, Strasbourg, France (DOI : 10.26093/cds/vizier). The original description of the VizieR service was published in 2000, A$\&$AS 143, 23

\section*{Data Availability}

The UVIT data used in this work is publicly available at \url{https://webapps.issdc.gov.in/astro_archive/archive/Home.jsp}. The other archival data used are available in the following links: GALEX: \url{https://mast.stsci.edu/portal/Mashup/Clients/Mast/Portal.html}, SDSS: \url{https://dr12.sdss.org/mosaics}, Pan-STARRS: \url{https://outerspace.stsci.edu/display/PANSTARRS/}, DECaLS: \url{https://www.legacysurvey.org/}, 2MASS, WISE and IRAS: \url{https://irsa.ipac.caltech.edu/frontpage/}. The softwares used for data analysis are available at, CCDLAB: \url{https://github.com/user29A/CCDLAB}, ASTRODENDRO: \url{https://dendrograms.readthedocs.io/en/stable/}, \url{https://github.com/dendrograms/astrodendro}, CIGALE: \url{https://cigale.lam.fr/}, \url{https://gitlab.lam.fr/cigale/cigale/}.



\bibliographystyle{mnras}
\bibliography{uvit} % if your bibtex file is called example.bib

\begin{thebibliography}{}
\makeatletter
\relax
\def\mn@urlcharsother{\let\do\@makeother \do\$\do\&\do\#\do\^\do\_\do\%\do\~}
\def\mn@doi{\begingroup\mn@urlcharsother \@ifnextchar [ {\mn@doi@} {\mn@doi@[]}}
\def\mn@doi@[#1]#2{\def\@tempa{#1}\ifx\@tempa\@empty \href {http://dx.doi.org/#2} {doi:#2}\else \href {http://dx.doi.org/#2} {#1}\fi \endgroup}
\def\mn@eprint#1#2{\mn@eprint@#1:#2::\@nil}
\def\mn@eprint@arXiv#1{\href {http://arxiv.org/abs/#1} {{\tt arXiv:#1}}}
\def\mn@eprint@dblp#1{\href {http://dblp.uni-trier.de/rec/bibtex/#1.xml} {dblp:#1}}
\def\mn@eprint@#1:#2:#3:#4\@nil{\def\@tempa {#1}\def\@tempb {#2}\def\@tempc {#3}\ifx \@tempc \@empty \let \@tempc \@tempb \let \@tempb \@tempa \fi \ifx \@tempb \@empty \def\@tempb {arXiv}\fi \@ifundefined {mn@eprint@\@tempb}{\@tempb:\@tempc}{\expandafter \expandafter \csname mn@eprint@\@tempb\endcsname \expandafter{\@tempc}}}

\bibitem[\protect\citeauthoryear{{Alexander} \& {Hickox}}{{Alexander} \& {Hickox}}{2012}]{2012NewAR..56...93A}
{Alexander} D.~M.,  {Hickox} R.~C.,  2012, \mn@doi [\nar] {10.1016/j.newar.2011.11.003}, \href {https://ui.adsabs.harvard.edu/abs/2012NewAR..56...93A} {56, 93}

\bibitem[\protect\citeauthoryear{{Alexander}, {Smail}, {Bauer}, {Chapman}, {Blain}, {Brandt}  \& {Ivison}}{{Alexander} et~al.}{2005}]{2005Natur.434..738A}
{Alexander} D.~M.,  {Smail} I.,  {Bauer} F.~E.,  {Chapman} S.~C.,  {Blain} A.~W.,  {Brandt} W.~N.,   {Ivison} R.~J.,  2005, \mn@doi [\nat] {10.1038/nature03473}, \href {https://ui.adsabs.harvard.edu/abs/2005Natur.434..738A} {434, 738}

\bibitem[\protect\citeauthoryear{{Bagchi} et~al.,}{{Bagchi} et~al.}{2014}]{2014ApJ...788..174B}
{Bagchi} J.,  et~al., 2014, \mn@doi [\apj] {10.1088/0004-637X/788/2/174}, \href {https://ui.adsabs.harvard.edu/abs/2014ApJ...788..174B} {788, 174}

\bibitem[\protect\citeauthoryear{{Behroozi}, {Conroy}  \& {Wechsler}}{{Behroozi} et~al.}{2010}]{2010ApJ...717..379B}
{Behroozi} P.~S.,  {Conroy} C.,   {Wechsler} R.~H.,  2010, \mn@doi [\apj] {10.1088/0004-637X/717/1/379}, \href {https://ui.adsabs.harvard.edu/abs/2010ApJ...717..379B} {717, 379}

\bibitem[\protect\citeauthoryear{{Behroozi}, {Wechsler}, {Hearin}  \& {Conroy}}{{Behroozi} et~al.}{2019}]{2019MNRAS.488.3143B}
{Behroozi} P.,  {Wechsler} R.~H.,  {Hearin} A.~P.,   {Conroy} C.,  2019, \mn@doi [\mnras] {10.1093/mnras/stz1182}, \href {https://ui.adsabs.harvard.edu/abs/2019MNRAS.488.3143B} {488, 3143}

\bibitem[\protect\citeauthoryear{{Bell}, {McIntosh}, {Katz}  \& {Weinberg}}{{Bell} et~al.}{2003}]{2003ApJS..149..289B}
{Bell} E.~F.,  {McIntosh} D.~H.,  {Katz} N.,   {Weinberg} M.~D.,  2003, \mn@doi [\apjs] {10.1086/378847}, \href {https://ui.adsabs.harvard.edu/abs/2003ApJS..149..289B} {149, 289}

\bibitem[\protect\citeauthoryear{{Benson}}{{Benson}}{2010}]{2010PhR...495...33B}
{Benson} A.~J.,  2010, \mn@doi [\physrep] {10.1016/j.physrep.2010.06.001}, \href {https://ui.adsabs.harvard.edu/abs/2010PhR...495...33B} {495, 33}

\bibitem[\protect\citeauthoryear{{Bertin}}{{Bertin}}{2010}]{2010ascl.soft10068B}
{Bertin} E.,  2010, {SWarp: Resampling and Co-adding FITS Images Together}, Astrophysics Source Code Library, record ascl:1010.068 (\mn@eprint {ascl} {1010.068})

\bibitem[\protect\citeauthoryear{{Bertin} \& {Arnouts}}{{Bertin} \& {Arnouts}}{1996}]{1996A&AS..117..393B}
{Bertin} E.,  {Arnouts} S.,  1996, \mn@doi [\aaps] {10.1051/aas:1996164}, \href {https://ui.adsabs.harvard.edu/abs/1996A&AS..117..393B} {117, 393}

\bibitem[\protect\citeauthoryear{{Bertin}, {Mellier}, {Radovich}, {Missonnier}, {Didelon}  \& {Morin}}{{Bertin} et~al.}{2002}]{2002ASPC..281..228B}
{Bertin} E.,  {Mellier} Y.,  {Radovich} M.,  {Missonnier} G.,  {Didelon} P.,   {Morin} B.,  2002, in {Bohlender} D.~A.,  {Durand} D.,   {Handley} T.~H.,  eds,  Astronomical Society of the Pacific Conference Series Vol. 281, Astronomical Data Analysis Software and Systems XI. p.~228

\bibitem[\protect\citeauthoryear{{Birnboim} \& {Dekel}}{{Birnboim} \& {Dekel}}{2003}]{2003MNRAS.345..349B}
{Birnboim} Y.,  {Dekel} A.,  2003, \mn@doi [\mnras] {10.1046/j.1365-8711.2003.06955.x}, \href {https://ui.adsabs.harvard.edu/abs/2003MNRAS.345..349B} {345, 349}

\bibitem[\protect\citeauthoryear{{Bogd{\'a}n} et~al.,}{{Bogd{\'a}n} et~al.}{2013a}]{2013ApJ...772...97B}
{Bogd{\'a}n} {\'A}.,  et~al., 2013a, \mn@doi [\apj] {10.1088/0004-637X/772/2/97}, \href {https://ui.adsabs.harvard.edu/abs/2013ApJ...772...97B} {772, 97}

\bibitem[\protect\citeauthoryear{{Bogd{\'a}n}, {Forman}, {Kraft}  \& {Jones}}{{Bogd{\'a}n} et~al.}{2013b}]{2013ApJ...772...98B}
{Bogd{\'a}n} {\'A}.,  {Forman} W.~R.,  {Kraft} R.~P.,   {Jones} C.,  2013b, \mn@doi [\apj] {10.1088/0004-637X/772/2/98}, \href {https://ui.adsabs.harvard.edu/abs/2013ApJ...772...98B} {772, 98}

\bibitem[\protect\citeauthoryear{{Boquien}, {Burgarella}, {Roehlly}, {Buat}, {Ciesla}, {Corre}, {Inoue}  \& {Salas}}{{Boquien} et~al.}{2019}]{2019A&A...622A.103B}
{Boquien} M.,  {Burgarella} D.,  {Roehlly} Y.,  {Buat} V.,  {Ciesla} L.,  {Corre} D.,  {Inoue} A.~K.,   {Salas} H.,  2019, \mn@doi [\aap] {10.1051/0004-6361/201834156}, \href {https://ui.adsabs.harvard.edu/abs/2019A&A...622A.103B} {622, A103}

\bibitem[\protect\citeauthoryear{{Bournaud}}{{Bournaud}}{2011}]{2011EAS....51..107B}
{Bournaud} F.,  2011, in {Charbonnel} C.,  {Montmerle} T.,  eds,  EAS Publications Series Vol. 51, EAS Publications Series. pp 107--131 (\mn@eprint {arXiv} {1106.1793}), \mn@doi{10.1051/eas/1151008}

\bibitem[\protect\citeauthoryear{{Bower}, {Benson}, {Malbon}, {Helly}, {Frenk}, {Baugh}, {Cole}  \& {Lacey}}{{Bower} et~al.}{2006}]{2006MNRAS.370..645B}
{Bower} R.~G.,  {Benson} A.~J.,  {Malbon} R.,  {Helly} J.~C.,  {Frenk} C.~S.,  {Baugh} C.~M.,  {Cole} S.,   {Lacey} C.~G.,  2006, \mn@doi [\mnras] {10.1111/j.1365-2966.2006.10519.x}, \href {https://ui.adsabs.harvard.edu/abs/2006MNRAS.370..645B} {370, 645}

\bibitem[\protect\citeauthoryear{{Brinchmann}, {Charlot}, {White}, {Tremonti}, {Kauffmann}, {Heckman}  \& {Brinkmann}}{{Brinchmann} et~al.}{2004}]{2004MNRAS.351.1151B}
{Brinchmann} J.,  {Charlot} S.,  {White} S.~D.~M.,  {Tremonti} C.,  {Kauffmann} G.,  {Heckman} T.,   {Brinkmann} J.,  2004, \mn@doi [\mnras] {10.1111/j.1365-2966.2004.07881.x}, \href {https://ui.adsabs.harvard.edu/abs/2004MNRAS.351.1151B} {351, 1151}

\bibitem[\protect\citeauthoryear{{Brodie} \& {Strader}}{{Brodie} \& {Strader}}{2006}]{2006ARA&A..44..193B}
{Brodie} J.~P.,  {Strader} J.,  2006, \mn@doi [\araa] {10.1146/annurev.astro.44.051905.092441}, \href {https://ui.adsabs.harvard.edu/abs/2006ARA&A..44..193B} {44, 193}

\bibitem[\protect\citeauthoryear{{Bruzual} \& {Charlot}}{{Bruzual} \& {Charlot}}{2003}]{2003MNRAS.344.1000B}
{Bruzual} G.,  {Charlot} S.,  2003, \mn@doi [\mnras] {10.1046/j.1365-8711.2003.06897.x}, \href {https://ui.adsabs.harvard.edu/abs/2003MNRAS.344.1000B} {344, 1000}

\bibitem[\protect\citeauthoryear{{Calzetti}, {Armus}, {Bohlin}, {Kinney}, {Koornneef}  \& {Storchi-Bergmann}}{{Calzetti} et~al.}{2000}]{2000ApJ...533..682C}
{Calzetti} D.,  {Armus} L.,  {Bohlin} R.~C.,  {Kinney} A.~L.,  {Koornneef} J.,   {Storchi-Bergmann} T.,  2000, \mn@doi [\apj] {10.1086/308692}, \href {https://ui.adsabs.harvard.edu/abs/2000ApJ...533..682C} {533, 682}

\bibitem[\protect\citeauthoryear{{Cardelli}, {Clayton}  \& {Mathis}}{{Cardelli} et~al.}{1989}]{1989ApJ...345..245C}
{Cardelli} J.~A.,  {Clayton} G.~C.,   {Mathis} J.~S.,  1989, \mn@doi [\apj] {10.1086/167900}, \href {https://ui.adsabs.harvard.edu/abs/1989ApJ...345..245C} {345, 245}

\bibitem[\protect\citeauthoryear{{Cheng} et~al.,}{{Cheng} et~al.}{2023}]{2023ApJ...942L..19C}
{Cheng} C.,  et~al., 2023, \mn@doi [\apjl] {10.3847/2041-8213/aca9d0}, \href {https://ui.adsabs.harvard.edu/abs/2023ApJ...942L..19C} {942, L19}

\bibitem[\protect\citeauthoryear{{Cicone} et~al.,}{{Cicone} et~al.}{2014}]{Ciconeetal2014}
{Cicone} C.,  et~al., 2014, \mn@doi [\aap] {10.1051/0004-6361/201322464}, \href {https://ui.adsabs.harvard.edu/abs/2014A&A...562A..21C} {562, A21}

\bibitem[\protect\citeauthoryear{{Cole}, {Lacey}, {Baugh}  \& {Frenk}}{{Cole} et~al.}{2000}]{2000MNRAS.319..168C}
{Cole} S.,  {Lacey} C.~G.,  {Baugh} C.~M.,   {Frenk} C.~S.,  2000, \mn@doi [\mnras] {10.1046/j.1365-8711.2000.03879.x}, \href {https://ui.adsabs.harvard.edu/abs/2000MNRAS.319..168C} {319, 168}

\bibitem[\protect\citeauthoryear{{Corradi} \& {Capaccioli}}{{Corradi} \& {Capaccioli}}{1991}]{1991A&AS...90..121C}
{Corradi} R.~L.~M.,  {Capaccioli} M.,  1991, \aaps, \href {https://ui.adsabs.harvard.edu/abs/1991A&AS...90..121C} {90, 121}

\bibitem[\protect\citeauthoryear{{Croton} et~al.,}{{Croton} et~al.}{2006}]{2006MNRAS.365...11C}
{Croton} D.~J.,  et~al., 2006, \mn@doi [\mnras] {10.1111/j.1365-2966.2005.09675.x}, \href {https://ui.adsabs.harvard.edu/abs/2006MNRAS.365...11C} {365, 11}

\bibitem[\protect\citeauthoryear{{Dai}, {Anderson}, {Bregman}  \& {Miller}}{{Dai} et~al.}{2012}]{2012ApJ...755..107D}
{Dai} X.,  {Anderson} M.~E.,  {Bregman} J.~N.,   {Miller} J.~M.,  2012, \mn@doi [\apj] {10.1088/0004-637X/755/2/107}, \href {https://ui.adsabs.harvard.edu/abs/2012ApJ...755..107D} {755, 107}

\bibitem[\protect\citeauthoryear{{Dekel}, {Sari}  \& {Ceverino}}{{Dekel} et~al.}{2009}]{2009ApJ...703..785D}
{Dekel} A.,  {Sari} R.,   {Ceverino} D.,  2009, \mn@doi [\apj] {10.1088/0004-637X/703/1/785}, \href {https://ui.adsabs.harvard.edu/abs/2009ApJ...703..785D} {703, 785}

\bibitem[\protect\citeauthoryear{{Dhiwar}, {Saha}, {Dekel}, {Paswan}, {Pandey}, {Cortesi}  \& {Pandge}}{{Dhiwar} et~al.}{2023}]{2023MNRAS.518.4943D}
{Dhiwar} S.,  {Saha} K.,  {Dekel} A.,  {Paswan} A.,  {Pandey} D.,  {Cortesi} A.,   {Pandge} M.,  2023, \mn@doi [\mnras] {10.1093/mnras/stac3369}, \href {https://ui.adsabs.harvard.edu/abs/2023MNRAS.518.4943D} {518, 4943}

\bibitem[\protect\citeauthoryear{{Draine} et~al.,}{{Draine} et~al.}{2014}]{2014ApJ...780..172D}
{Draine} B.~T.,  et~al., 2014, \mn@doi [\apj] {10.1088/0004-637X/780/2/172}, \href {https://ui.adsabs.harvard.edu/abs/2014ApJ...780..172D} {780, 172}

\bibitem[\protect\citeauthoryear{{Elbaz} et~al.,}{{Elbaz} et~al.}{2007}]{2007A&A...468...33E}
{Elbaz} D.,  et~al., 2007, \mn@doi [\aap] {10.1051/0004-6361:20077525}, \href {https://ui.adsabs.harvard.edu/abs/2007A&A...468...33E} {468, 33}

\bibitem[\protect\citeauthoryear{{Feldmann} \& {Mayer}}{{Feldmann} \& {Mayer}}{2015}]{2015MNRAS.446.1939F}
{Feldmann} R.,  {Mayer} L.,  2015, \mn@doi [\mnras] {10.1093/mnras/stu2207}, \href {https://ui.adsabs.harvard.edu/abs/2015MNRAS.446.1939F} {446, 1939}

\bibitem[\protect\citeauthoryear{{F{\"o}rster Schreiber} et~al.,}{{F{\"o}rster Schreiber} et~al.}{2014}]{ForsterSchreiberetal2014}
{F{\"o}rster Schreiber} N.~M.,  et~al., 2014, \mn@doi [\apj] {10.1088/0004-637X/787/1/38}, \href {https://ui.adsabs.harvard.edu/abs/2014ApJ...787...38F} {787, 38}

\bibitem[\protect\citeauthoryear{{Fritz}, {Franceschini}  \& {Hatziminaoglou}}{{Fritz} et~al.}{2006}]{2006MNRAS.366..767F}
{Fritz} J.,  {Franceschini} A.,   {Hatziminaoglou} E.,  2006, \mn@doi [\mnras] {10.1111/j.1365-2966.2006.09866.x}, \href {https://ui.adsabs.harvard.edu/abs/2006MNRAS.366..767F} {366, 767}

\bibitem[\protect\citeauthoryear{{Fukugita} \& {Peebles}}{{Fukugita} \& {Peebles}}{2006}]{2006ApJ...639..590F}
{Fukugita} M.,  {Peebles} P.~J.~E.,  2006, \mn@doi [\apj] {10.1086/499556}, \href {https://ui.adsabs.harvard.edu/abs/2006ApJ...639..590F} {639, 590}

\bibitem[\protect\citeauthoryear{{Garrido}, {Marcelin}, {Amram}, {Balkowski}, {Gach}  \& {Boulesteix}}{{Garrido} et~al.}{2005}]{2005MNRAS.362..127G}
{Garrido} O.,  {Marcelin} M.,  {Amram} P.,  {Balkowski} C.,  {Gach} J.~L.,   {Boulesteix} J.,  2005, \mn@doi [\mnras] {10.1111/j.1365-2966.2005.09274.x}, \href {https://ui.adsabs.harvard.edu/abs/2005MNRAS.362..127G} {362, 127}

\bibitem[\protect\citeauthoryear{{Gebhardt} et~al.,}{{Gebhardt} et~al.}{2000}]{2000ApJ...539L..13G}
{Gebhardt} K.,  et~al., 2000, \mn@doi [\apjl] {10.1086/312840}, \href {https://ui.adsabs.harvard.edu/abs/2000ApJ...539L..13G} {539, L13}

\bibitem[\protect\citeauthoryear{{Gil de Paz} et~al.,}{{Gil de Paz} et~al.}{2007}]{2007ApJS..173..185G}
{Gil de Paz} A.,  et~al., 2007, \mn@doi [\apjs] {10.1086/516636}, \href {https://ui.adsabs.harvard.edu/abs/2007ApJS..173..185G} {173, 185}

\bibitem[\protect\citeauthoryear{{Gott}}{{Gott}}{1977}]{1977ARA&A..15..235G}
{Gott} J.~R. I.,  1977, \mn@doi [\araa] {10.1146/annurev.aa.15.090177.001315}, \href {https://ui.adsabs.harvard.edu/abs/1977ARA&A..15..235G} {15, 235}

\bibitem[\protect\citeauthoryear{{Gottesman}, {Hunter}  \& {Shostak}}{{Gottesman} et~al.}{1983}]{1983MNRAS.202P..21G}
{Gottesman} S.~T.,  {Hunter} J.~H. J.,   {Shostak} G.~S.,  1983, \mn@doi [\mnras] {10.1093/mnras/202.1.21P}, \href {https://ui.adsabs.harvard.edu/abs/1983MNRAS.202P..21G} {202, 21P}

\bibitem[\protect\citeauthoryear{{G{\"u}ltekin} et~al.,}{{G{\"u}ltekin} et~al.}{2009}]{2009ApJ...698..198G}
{G{\"u}ltekin} K.,  et~al., 2009, \mn@doi [\apj] {10.1088/0004-637X/698/1/198}, \href {https://ui.adsabs.harvard.edu/abs/2009ApJ...698..198G} {698, 198}

\bibitem[\protect\citeauthoryear{{Haynes} et~al.,}{{Haynes} et~al.}{2011}]{2011AJ....142..170H}
{Haynes} M.~P.,  et~al., 2011, \mn@doi [\aj] {10.1088/0004-6256/142/5/170}, \href {https://ui.adsabs.harvard.edu/abs/2011AJ....142..170H} {142, 170}

\bibitem[\protect\citeauthoryear{{Helou}, {Khan}, {Malek}  \& {Boehmer}}{{Helou} et~al.}{1988}]{1988ApJS...68..151H}
{Helou} G.,  {Khan} I.~R.,  {Malek} L.,   {Boehmer} L.,  1988, \mn@doi [\apjs] {10.1086/191285}, \href {https://ui.adsabs.harvard.edu/abs/1988ApJS...68..151H} {68, 151}

\bibitem[\protect\citeauthoryear{{Hibbard}, {van Gorkom}, {Rupen}  \& {Schiminovich}}{{Hibbard} et~al.}{2001}]{2001ASPC..240..657H}
{Hibbard} J.~E.,  {van Gorkom} J.~H.,  {Rupen} M.~P.,   {Schiminovich} D.,  2001, in {Hibbard} J.~E.,  {Rupen} M.,   {van Gorkom} J.~H.,  eds,  Astronomical Society of the Pacific Conference Series Vol. 240, Gas and Galaxy Evolution. p.~657 (\mn@eprint {arXiv} {astro-ph/0110667}), \mn@doi{10.48550/arXiv.astro-ph/0110667}

\bibitem[\protect\citeauthoryear{{Hoopes} et~al.,}{{Hoopes} et~al.}{2007}]{2007ApJS..173..441H}
{Hoopes} C.~G.,  et~al., 2007, \mn@doi [\apjs] {10.1086/516644}, \href {https://ui.adsabs.harvard.edu/abs/2007ApJS..173..441H} {173, 441}

\bibitem[\protect\citeauthoryear{Hopkins, Narayan  \& Hernquist}{Hopkins et~al.}{2006}]{Hopkins_2006}
Hopkins P.~F.,  Narayan R.,   Hernquist L.,  2006, \mn@doi [The Astrophysical Journal] {10.1086/503154}, 643, 641

\bibitem[\protect\citeauthoryear{{Huang} et~al.,}{{Huang} et~al.}{2023}]{2023arXiv230401378H}
{Huang} J.~S.,  et~al., 2023, \mn@doi [arXiv e-prints] {10.48550/arXiv.2304.01378}, \href {https://ui.adsabs.harvard.edu/abs/2023arXiv230401378H} {p. arXiv:2304.01378}

\bibitem[\protect\citeauthoryear{{Inoue}}{{Inoue}}{2011}]{2011MNRAS.415.2920I}
{Inoue} A.~K.,  2011, \mn@doi [\mnras] {10.1111/j.1365-2966.2011.18906.x}, \href {https://ui.adsabs.harvard.edu/abs/2011MNRAS.415.2920I} {415, 2920}

\bibitem[\protect\citeauthoryear{{Joye} \& {Mandel}}{{Joye} \& {Mandel}}{2003}]{2003ASPC..295..489J}
{Joye} W.~A.,  {Mandel} E.,  2003, in {Payne} H.~E.,  {Jedrzejewski} R.~I.,   {Hook} R.~N.,  eds,  Astronomical Society of the Pacific Conference Series Vol. 295, Astronomical Data Analysis Software and Systems XII. p.~489

\bibitem[\protect\citeauthoryear{{Kalinova}, {Colombo}, {S{\'a}nchez}, {Kodaira}, {Garc{\'\i}a-Benito}, {Gonz{\'a}lez Delgado}, {Rosolowsky}  \& {Lacerda}}{{Kalinova} et~al.}{2021}]{2021A&A...648A..64K}
{Kalinova} V.,  {Colombo} D.,  {S{\'a}nchez} S.~F.,  {Kodaira} K.,  {Garc{\'\i}a-Benito} R.,  {Gonz{\'a}lez Delgado} R.,  {Rosolowsky} E.,   {Lacerda} E.~A.~D.,  2021, \mn@doi [\aap] {10.1051/0004-6361/202039896}, \href {https://ui.adsabs.harvard.edu/abs/2021A&A...648A..64K} {648, A64}

\bibitem[\protect\citeauthoryear{{Kaviraj} et~al.,}{{Kaviraj} et~al.}{2007}]{2007ApJS..173..619K}
{Kaviraj} S.,  et~al., 2007, \mn@doi [\apjs] {10.1086/516633}, \href {https://ui.adsabs.harvard.edu/abs/2007ApJS..173..619K} {173, 619}

\bibitem[\protect\citeauthoryear{{Kelly}, {Jenkins}  \& {Frenk}}{{Kelly} et~al.}{2021}]{2021MNRAS.502.2934K}
{Kelly} A.~J.,  {Jenkins} A.,   {Frenk} C.~S.,  2021, \mn@doi [\mnras] {10.1093/mnras/stab255}, \href {https://ui.adsabs.harvard.edu/abs/2021MNRAS.502.2934K} {502, 2934}

\bibitem[\protect\citeauthoryear{{Kennicutt}}{{Kennicutt}}{1998}]{1998ARA&A..36..189K}
{Kennicutt} Robert~C. J.,  1998, \mn@doi [\araa] {10.1146/annurev.astro.36.1.189}, \href {https://ui.adsabs.harvard.edu/abs/1998ARA&A..36..189K} {36, 189}

\bibitem[\protect\citeauthoryear{{Kennicutt} \& {Evans}}{{Kennicutt} \& {Evans}}{2012}]{2012ARA&A..50..531K}
{Kennicutt} R.~C.,  {Evans} N.~J.,  2012, \mn@doi [\araa] {10.1146/annurev-astro-081811-125610}, \href {https://ui.adsabs.harvard.edu/abs/2012ARA&A..50..531K} {50, 531}

\bibitem[\protect\citeauthoryear{{Komatsu} et~al.,}{{Komatsu} et~al.}{2011}]{2011ApJS..192...18K}
{Komatsu} E.,  et~al., 2011, \mn@doi [\apjs] {10.1088/0067-0049/192/2/18}, \href {https://ui.adsabs.harvard.edu/abs/2011ApJS..192...18K} {192, 18}

\bibitem[\protect\citeauthoryear{{Labbe} et~al.,}{{Labbe} et~al.}{2022}]{2022arXiv220712446L}
{Labbe} I.,  et~al., 2022, \mn@doi [arXiv e-prints] {10.48550/arXiv.2207.12446}, \href {https://ui.adsabs.harvard.edu/abs/2022arXiv220712446L} {p. arXiv:2207.12446}

\bibitem[\protect\citeauthoryear{{Lackner}, {Cen}, {Ostriker}  \& {Joung}}{{Lackner} et~al.}{2012}]{2012MNRAS.425..641L}
{Lackner} C.~N.,  {Cen} R.,  {Ostriker} J.~P.,   {Joung} M.~R.,  2012, \mn@doi [\mnras] {10.1111/j.1365-2966.2012.21525.x}, \href {https://ui.adsabs.harvard.edu/abs/2012MNRAS.425..641L} {425, 641}

\bibitem[\protect\citeauthoryear{{Lelli}, {McGaugh}  \& {Schombert}}{{Lelli} et~al.}{2016}]{2016AJ....152..157L}
{Lelli} F.,  {McGaugh} S.~S.,   {Schombert} J.~M.,  2016, \mn@doi [\aj] {10.3847/0004-6256/152/6/157}, \href {https://ui.adsabs.harvard.edu/abs/2016AJ....152..157L} {152, 157}

\bibitem[\protect\citeauthoryear{{L{\"u}tticke}, {Pohlen}  \& {Dettmar}}{{L{\"u}tticke} et~al.}{2004}]{2004A&A...417..527L}
{L{\"u}tticke} R.,  {Pohlen} M.,   {Dettmar} R.~J.,  2004, \mn@doi [\aap] {10.1051/0004-6361:20031782}, \href {https://ui.adsabs.harvard.edu/abs/2004A&A...417..527L} {417, 527}

\bibitem[\protect\citeauthoryear{{Madau} \& {Dickinson}}{{Madau} \& {Dickinson}}{2014}]{MadauDickinson2014}
{Madau} P.,  {Dickinson} M.,  2014, \mn@doi [\araa] {10.1146/annurev-astro-081811-125615}, \href {https://ui.adsabs.harvard.edu/abs/2014ARA&A..52..415M} {52, 415}

\bibitem[\protect\citeauthoryear{{Magorrian} et~al.,}{{Magorrian} et~al.}{1998}]{1998AJ....115.2285M}
{Magorrian} J.,  et~al., 1998, \mn@doi [\aj] {10.1086/300353}, \href {https://ui.adsabs.harvard.edu/abs/1998AJ....115.2285M} {115, 2285}

\bibitem[\protect\citeauthoryear{{Makarov}, {Prugniel}, {Terekhova}, {Courtois}  \& {Vauglin}}{{Makarov} et~al.}{2014}]{2014A&A...570A..13M}
{Makarov} D.,  {Prugniel} P.,  {Terekhova} N.,  {Courtois} H.,   {Vauglin} I.,  2014, \mn@doi [\aap] {10.1051/0004-6361/201423496}, \href {http://adsabs.harvard.edu/abs/2014A%26A...570A..13M} {570, A13}

\bibitem[\protect\citeauthoryear{{Man} \& {Belli}}{{Man} \& {Belli}}{2018}]{2018NatAs...2..695M}
{Man} A.,  {Belli} S.,  2018, \mn@doi [Nature Astronomy] {10.1038/s41550-018-0558-1}, \href {https://ui.adsabs.harvard.edu/abs/2018NatAs...2..695M} {2, 695}

\bibitem[\protect\citeauthoryear{{Mart{\'\i}n-Navarro}, {Brodie}, {Romanowsky}, {Ruiz-Lara}  \& {van de Ven}}{{Mart{\'\i}n-Navarro} et~al.}{2018}]{2018Natur.553..307M}
{Mart{\'\i}n-Navarro} I.,  {Brodie} J.~P.,  {Romanowsky} A.~J.,  {Ruiz-Lara} T.,   {van de Ven} G.,  2018, \mn@doi [\nat] {10.1038/nature24999}, \href {https://ui.adsabs.harvard.edu/abs/2018Natur.553..307M} {553, 307}

\bibitem[\protect\citeauthoryear{{Mathis}, {Rumpl}  \& {Nordsieck}}{{Mathis} et~al.}{1977}]{1977ApJ...217..425M}
{Mathis} J.~S.,  {Rumpl} W.,   {Nordsieck} K.~H.,  1977, \mn@doi [\apj] {10.1086/155591}, \href {https://ui.adsabs.harvard.edu/abs/1977ApJ...217..425M} {217, 425}

\bibitem[\protect\citeauthoryear{{McGaugh}}{{McGaugh}}{2005}]{2005ApJ...632..859M}
{McGaugh} S.~S.,  2005, \mn@doi [\apj] {10.1086/432968}, \href {https://ui.adsabs.harvard.edu/abs/2005ApJ...632..859M} {632, 859}

\bibitem[\protect\citeauthoryear{{Mirakhor} et~al.,}{{Mirakhor} et~al.}{2021}]{2021MNRAS.500.2503M}
{Mirakhor} M.~S.,  et~al., 2021, \mn@doi [\mnras] {10.1093/mnras/staa3404}, \href {https://ui.adsabs.harvard.edu/abs/2021MNRAS.500.2503M} {500, 2503}

\bibitem[\protect\citeauthoryear{{Moster}, {Naab}  \& {White}}{{Moster} et~al.}{2013}]{2013MNRAS.428.3121M}
{Moster} B.~P.,  {Naab} T.,   {White} S. D.~M.,  2013, \mn@doi [\mnras] {10.1093/mnras/sts261}, \href {https://ui.adsabs.harvard.edu/abs/2013MNRAS.428.3121M} {428, 3121}

\bibitem[\protect\citeauthoryear{{Naab} \& {Ostriker}}{{Naab} \& {Ostriker}}{2017}]{2017ARA&A..55...59N}
{Naab} T.,  {Ostriker} J.~P.,  2017, \mn@doi [\araa] {10.1146/annurev-astro-081913-040019}, \href {https://ui.adsabs.harvard.edu/abs/2017ARA&A..55...59N} {55, 59}

\bibitem[\protect\citeauthoryear{{Nesvadba} et~al.,}{{Nesvadba} et~al.}{2021}]{2021A&A...654A...8N}
{Nesvadba} N.~P.~H.,  et~al., 2021, \mn@doi [\aap] {10.1051/0004-6361/202140544}, \href {https://ui.adsabs.harvard.edu/abs/2021A&A...654A...8N} {654, A8}

\bibitem[\protect\citeauthoryear{Ogle, Lanz, Appleton, Helou  \& Mazzarella}{Ogle et~al.}{2019a}]{Ogle_2019}
Ogle P.~M.,  Lanz L.,  Appleton P.~N.,  Helou G.,   Mazzarella J.,  2019a, \mn@doi [The Astrophysical Journal Supplement Series] {10.3847/1538-4365/ab21c3}, 243, 14

\bibitem[\protect\citeauthoryear{{Ogle}, {Jarrett}, {Lanz}, {Cluver}, {Alatalo}, {Appleton}  \& {Mazzarella}}{{Ogle} et~al.}{2019b}]{2019ApJ...884L..11O}
{Ogle} P.~M.,  {Jarrett} T.,  {Lanz} L.,  {Cluver} M.,  {Alatalo} K.,  {Appleton} P.~N.,   {Mazzarella} J.~M.,  2019b, \mn@doi [\apjl] {10.3847/2041-8213/ab459e}, \href {https://ui.adsabs.harvard.edu/abs/2019ApJ...884L..11O} {884, L11}

\bibitem[\protect\citeauthoryear{{Oser}, {Ostriker}, {Naab}, {Johansson}  \& {Burkert}}{{Oser} et~al.}{2010}]{2010ApJ...725.2312O}
{Oser} L.,  {Ostriker} J.~P.,  {Naab} T.,  {Johansson} P.~H.,   {Burkert} A.,  2010, \mn@doi [\apj] {10.1088/0004-637X/725/2/2312}, \href {https://ui.adsabs.harvard.edu/abs/2010ApJ...725.2312O} {725, 2312}

\bibitem[\protect\citeauthoryear{{Park} et~al.,}{{Park} et~al.}{2023}]{2023ApJ...953..119P}
{Park} M.,  et~al., 2023, \mn@doi [\apj] {10.3847/1538-4357/acd54a}, \href {https://ui.adsabs.harvard.edu/abs/2023ApJ...953..119P} {953, 119}

\bibitem[\protect\citeauthoryear{{Pearson} et~al.,}{{Pearson} et~al.}{2019}]{2019A&A...631A..51P}
{Pearson} W.~J.,  et~al., 2019, \mn@doi [\aap] {10.1051/0004-6361/201936337}, \href {https://ui.adsabs.harvard.edu/abs/2019A&A...631A..51P} {631, A51}

\bibitem[\protect\citeauthoryear{{Posti}, {Fraternali}  \& {Marasco}}{{Posti} et~al.}{2019}]{2019A&A...626A..56P}
{Posti} L.,  {Fraternali} F.,   {Marasco} A.,  2019, \mn@doi [\aap] {10.1051/0004-6361/201935553}, \href {https://ui.adsabs.harvard.edu/abs/2019A&A...626A..56P} {626, A56}

\bibitem[\protect\citeauthoryear{{Postma} \& {Leahy}}{{Postma} \& {Leahy}}{2017}]{2017PASP..129k5002P}
{Postma} J.~E.,  {Leahy} D.,  2017, \mn@doi [\pasp] {10.1088/1538-3873/aa8800}, \href {https://ui.adsabs.harvard.edu/abs/2017PASP..129k5002P} {129, 115002}

\bibitem[\protect\citeauthoryear{{Ray}, {Bagchi}, {Dhiwar}, {Pandge}, {Mirakhor}, {Walker}  \& {Mukherjee}}{{Ray} et~al.}{2022}]{2022MNRAS.517...99R}
{Ray} S.,  {Bagchi} J.,  {Dhiwar} S.,  {Pandge} M.~B.,  {Mirakhor} M.,  {Walker} S.~A.,   {Mukherjee} D.,  2022, \mn@doi [\mnras] {10.1093/mnras/stac2683}, \href {https://ui.adsabs.harvard.edu/abs/2022MNRAS.517...99R} {517, 99}

\bibitem[\protect\citeauthoryear{{Rees} \& {Ostriker}}{{Rees} \& {Ostriker}}{1977}]{1977MNRAS.179..541R}
{Rees} M.~J.,  {Ostriker} J.~P.,  1977, \mn@doi [\mnras] {10.1093/mnras/179.4.541}, \href {https://ui.adsabs.harvard.edu/abs/1977MNRAS.179..541R} {179, 541}

\bibitem[\protect\citeauthoryear{{Reines} \& {Volonteri}}{{Reines} \& {Volonteri}}{2015}]{2015ApJ...813...82R}
{Reines} A.~E.,  {Volonteri} M.,  2015, \mn@doi [\apj] {10.1088/0004-637X/813/2/82}, \href {https://ui.adsabs.harvard.edu/abs/2015ApJ...813...82R} {813, 82}

\bibitem[\protect\citeauthoryear{{Renzini} \& {Peng}}{{Renzini} \& {Peng}}{2015}]{2015ApJ...801L..29R}
{Renzini} A.,  {Peng} Y.-j.,  2015, \mn@doi [\apjl] {10.1088/2041-8205/801/2/L29}, \href {https://ui.adsabs.harvard.edu/abs/2015ApJ...801L..29R} {801, L29}

\bibitem[\protect\citeauthoryear{{Robertson}}{{Robertson}}{2022}]{2022ARA&A..60..121R}
{Robertson} B.~E.,  2022, \mn@doi [\araa] {10.1146/annurev-astro-120221-044656}, \href {https://ui.adsabs.harvard.edu/abs/2022ARA&A..60..121R} {60, 121}

\bibitem[\protect\citeauthoryear{{Robitaille}, {Rice}, {Beaumont}, {Ginsburg}, {MacDonald}  \& {Rosolowsky}}{{Robitaille} et~al.}{2019}]{2019ascl.soft07016R}
{Robitaille} T.,  {Rice} T.,  {Beaumont} C.,  {Ginsburg} A.,  {MacDonald} B.,   {Rosolowsky} E.,  2019, {astrodendro: Astronomical data dendrogram creator}, Astrophysics Source Code Library, record ascl:1907.016 (\mn@eprint {ascl} {1907.016})

\bibitem[\protect\citeauthoryear{{Rodriguez-Gomez} et~al.,}{{Rodriguez-Gomez} et~al.}{2016}]{2016MNRAS.458.2371R}
{Rodriguez-Gomez} V.,  et~al., 2016, \mn@doi [\mnras] {10.1093/mnras/stw456}, \href {https://ui.adsabs.harvard.edu/abs/2016MNRAS.458.2371R} {458, 2371}

\bibitem[\protect\citeauthoryear{{Rubin}, {Ford}  \& {Roberts}}{{Rubin} et~al.}{1979}]{1979ApJ...230...35R}
{Rubin} V.~C.,  {Ford} W.~K.~J.,   {Roberts} M.~S.,  1979, \mn@doi [\apj] {10.1086/157059}, \href {https://ui.adsabs.harvard.edu/abs/1979ApJ...230...35R} {230, 35}

\bibitem[\protect\citeauthoryear{{Saglia} \& {Sancisi}}{{Saglia} \& {Sancisi}}{1988}]{1988A&A...203...28S}
{Saglia} R.~P.,  {Sancisi} R.,  1988, \aap, \href {https://ui.adsabs.harvard.edu/abs/1988A&A...203...28S} {203, 28}

\bibitem[\protect\citeauthoryear{{Saha}, {Dhiwar}, {Barway}, {Narayan}  \& {Tandon}}{{Saha} et~al.}{2021}]{2021JApA...42...59S}
{Saha} K.,  {Dhiwar} S.,  {Barway} S.,  {Narayan} C.,   {Tandon} S.,  2021, \mn@doi [Journal of Astrophysics and Astronomy] {10.1007/s12036-021-09715-5}, \href {https://ui.adsabs.harvard.edu/abs/2021JApA...42...59S} {42, 59}

\bibitem[\protect\citeauthoryear{{Salim} et~al.,}{{Salim} et~al.}{2016}]{2016ApJS..227....2S}
{Salim} S.,  et~al., 2016, \mn@doi [\apjs] {10.3847/0067-0049/227/1/2}, \href {https://ui.adsabs.harvard.edu/abs/2016ApJS..227....2S} {227, 2}

\bibitem[\protect\citeauthoryear{{Salpeter}}{{Salpeter}}{1955}]{1955ApJ...121..161S}
{Salpeter} E.~E.,  1955, \mn@doi [\apj] {10.1086/145971}, \href {https://ui.adsabs.harvard.edu/abs/1955ApJ...121..161S} {121, 161}

\bibitem[\protect\citeauthoryear{{Schlafly} \& {Finkbeiner}}{{Schlafly} \& {Finkbeiner}}{2011}]{2011ApJ...737..103S}
{Schlafly} E.~F.,  {Finkbeiner} D.~P.,  2011, \mn@doi [\apj] {10.1088/0004-637X/737/2/103}, \href {https://ui.adsabs.harvard.edu/abs/2011ApJ...737..103S} {737, 103}

\bibitem[\protect\citeauthoryear{{Schultz} \& {Wiemer}}{{Schultz} \& {Wiemer}}{1975}]{1975A&A....43..133S}
{Schultz} G.~V.,  {Wiemer} W.,  1975, \aap, \href {https://ui.adsabs.harvard.edu/abs/1975A&A....43..133S} {43, 133}

\bibitem[\protect\citeauthoryear{Shapiro}{Shapiro}{2005}]{Shapiro_2005}
Shapiro S.~L.,  2005, \mn@doi [The Astrophysical Journal] {10.1086/427065}, 620, 59

\bibitem[\protect\citeauthoryear{{Shaver}, {Hook}, {Jackson}, {Wall}  \& {Kellermann}}{{Shaver} et~al.}{1999}]{1999ASPC..156..163S}
{Shaver} P.~A.,  {Hook} I.~M.,  {Jackson} C.~A.,  {Wall} J.~V.,   {Kellermann} K.~I.,  1999, in {Carilli} C.~L.,  {Radford} S.~J.~E.,  {Menten} K.~M.,   {Langston} G.~I.,  eds,  Astronomical Society of the Pacific Conference Series Vol. 156, Highly Redshifted Radio Lines. p.~163 (\mn@eprint {arXiv} {astro-ph/9801211}), \mn@doi{10.48550/arXiv.astro-ph/9801211}

\bibitem[\protect\citeauthoryear{{Shostak}, {Hummel}, {Shaver}, {van der Hulst}  \& {van der Kruit}}{{Shostak} et~al.}{1982}]{1982A&A...115..293S}
{Shostak} G.~S.,  {Hummel} E.,  {Shaver} P.~A.,  {van der Hulst} J.~M.,   {van der Kruit} P.~C.,  1982, \aap, \href {https://ui.adsabs.harvard.edu/abs/1982A&A...115..293S} {115, 293}

\bibitem[\protect\citeauthoryear{{Somerville} \& {Dav{\'e}}}{{Somerville} \& {Dav{\'e}}}{2015}]{2015ARA&A..53...51S}
{Somerville} R.~S.,  {Dav{\'e}} R.,  2015, \mn@doi [\araa] {10.1146/annurev-astro-082812-140951}, \href {https://ui.adsabs.harvard.edu/abs/2015ARA&A..53...51S} {53, 51}

\bibitem[\protect\citeauthoryear{{Sommer-Larsen}}{{Sommer-Larsen}}{2006}]{2006ApJ...644L...1S}
{Sommer-Larsen} J.,  2006, \mn@doi [\apjl] {10.1086/505489}, \href {https://ui.adsabs.harvard.edu/abs/2006ApJ...644L...1S} {644, L1}

\bibitem[\protect\citeauthoryear{{Springob}, {Haynes}, {Giovanelli}  \& {Kent}}{{Springob} et~al.}{2005}]{2005ApJS..160..149S}
{Springob} C.~M.,  {Haynes} M.~P.,  {Giovanelli} R.,   {Kent} B.~R.,  2005, \mn@doi [\apjs] {10.1086/431550}, \href {https://ui.adsabs.harvard.edu/abs/2005ApJS..160..149S} {160, 149}

\bibitem[\protect\citeauthoryear{{Tandon} et~al.,}{{Tandon} et~al.}{2017}]{2017AJ....154..128T}
{Tandon} S.~N.,  et~al., 2017, \mn@doi [\aj] {10.3847/1538-3881/aa8451}, \href {https://ui.adsabs.harvard.edu/abs/2017AJ....154..128T} {154, 128}

\bibitem[\protect\citeauthoryear{{Thomas}, {Maraston}, {Bender}  \& {Mendes de Oliveira}}{{Thomas} et~al.}{2005}]{2005ApJ...621..673T}
{Thomas} D.,  {Maraston} C.,  {Bender} R.,   {Mendes de Oliveira} C.,  2005, \mn@doi [\apj] {10.1086/426932}, \href {https://ui.adsabs.harvard.edu/abs/2005ApJ...621..673T} {621, 673}

\bibitem[\protect\citeauthoryear{{Thomas}, {Maraston}, {Schawinski}, {Sarzi}  \& {Silk}}{{Thomas} et~al.}{2010}]{2010MNRAS.404.1775T}
{Thomas} D.,  {Maraston} C.,  {Schawinski} K.,  {Sarzi} M.,   {Silk} J.,  2010, \mn@doi [\mnras] {10.1111/j.1365-2966.2010.16427.x}, \href {https://ui.adsabs.harvard.edu/abs/2010MNRAS.404.1775T} {404, 1775}

\bibitem[\protect\citeauthoryear{{Tifft} \& {Cocke}}{{Tifft} \& {Cocke}}{1988}]{1988ApJS...67....1T}
{Tifft} W.~G.,  {Cocke} W.~J.,  1988, \mn@doi [\apjs] {10.1086/191265}, \href {https://ui.adsabs.harvard.edu/abs/1988ApJS...67....1T} {67, 1}

\bibitem[\protect\citeauthoryear{{Tody}}{{Tody}}{1986}]{1986SPIE..627..733T}
{Tody} D.,  1986, in {Crawford} D.~L.,  ed.,  Society of Photo-Optical Instrumentation Engineers (SPIE) Conference Series Vol. 627, Instrumentation in astronomy VI. p.~733, \mn@doi{10.1117/12.968154}

\bibitem[\protect\citeauthoryear{{Vollmer}, {Soida}, {Chung}, {van Gorkom}, {Otmianowska-Mazur}, {Beck}, {Urbanik}  \& {Kenney}}{{Vollmer} et~al.}{2008}]{2008A&A...483...89V}
{Vollmer} B.,  {Soida} M.,  {Chung} A.,  {van Gorkom} J.~H.,  {Otmianowska-Mazur} K.,  {Beck} R.,  {Urbanik} M.,   {Kenney} J.~D.~P.,  2008, \mn@doi [\aap] {10.1051/0004-6361:20078139}, \href {https://ui.adsabs.harvard.edu/abs/2008A&A...483...89V} {483, 89}

\bibitem[\protect\citeauthoryear{{Wechsler} \& {Tinker}}{{Wechsler} \& {Tinker}}{2018}]{2018ARA&A..56..435W}
{Wechsler} R.~H.,  {Tinker} J.~L.,  2018, \mn@doi [\araa] {10.1146/annurev-astro-081817-051756}, \href {https://ui.adsabs.harvard.edu/abs/2018ARA&A..56..435W} {56, 435}

\bibitem[\protect\citeauthoryear{{White} \& {Frenk}}{{White} \& {Frenk}}{1991}]{1991ApJ...379...52W}
{White} S. D.~M.,  {Frenk} C.~S.,  1991, \mn@doi [\apj] {10.1086/170483}, \href {https://ui.adsabs.harvard.edu/abs/1991ApJ...379...52W} {379, 52}

\bibitem[\protect\citeauthoryear{{White} \& {Rees}}{{White} \& {Rees}}{1978}]{1978MNRAS.183..341W}
{White} S.~D.~M.,  {Rees} M.~J.,  1978, \mn@doi [\mnras] {10.1093/mnras/183.3.341}, \href {https://ui.adsabs.harvard.edu/abs/1978MNRAS.183..341W} {183, 341}

\bibitem[\protect\citeauthoryear{{Xu}, {Liu}, {Jing}, {Wang}  \& {Lu}}{{Xu} et~al.}{2020}]{2020ApJ...895..100X}
{Xu} K.,  {Liu} C.,  {Jing} Y.,  {Wang} Y.,   {Lu} S.,  2020, \mn@doi [\apj] {10.3847/1538-4357/ab8fa0}, \href {https://ui.adsabs.harvard.edu/abs/2020ApJ...895..100X} {895, 100}

\bibitem[\protect\citeauthoryear{{Yang} et~al.,}{{Yang} et~al.}{2022}]{2022ApJ...927..192Y}
{Yang} G.,  et~al., 2022, \mn@doi [\apj] {10.3847/1538-4357/ac4971}, \href {https://ui.adsabs.harvard.edu/abs/2022ApJ...927..192Y} {927, 192}

\bibitem[\protect\citeauthoryear{{Young} \& {Knezek}}{{Young} \& {Knezek}}{1989}]{1989ApJ...347L..55Y}
{Young} J.~S.,  {Knezek} P.~M.,  1989, \mn@doi [\apjl] {10.1086/185606}, \href {https://ui.adsabs.harvard.edu/abs/1989ApJ...347L..55Y} {347, L55}

\makeatother
\end{thebibliography}

%\begin{thebibliography}{99}
%\end{thebibliography}

%%%%%%%%%%%%%%%%%%%%%%%%%%%%%%%%%%%%%%%%%%%%%%%%%%

%%%%%%%%%%%%%%%%% APPENDICES %%%%%%%%%%%%%%%%%%%%%

%\appendix



%%%%%%%%%%%%%%%%%%%%%%%%%%%%%%%%%%%%%%%%%%%%%%%%%%

% Don't change these lines
\bsp	% typesetting comment
\label{lastpage}
\end{document}